\documentclass[aps,preprintnumbers,superscriptaddress, unsortedaddress  ,nofootinbib, twocolumn]{revtex4}

\usepackage{ucs}
\usepackage{graphicx}
\usepackage{amssymb} 
\usepackage{amsmath} 
\usepackage{amsthm}
\usepackage[usenames,dvipsnames]{color}
\usepackage{bbm} % 1|
\usepackage{url}
\usepackage{natbib}
    \newcommand{\ket}[1]{\vert  #1 \rangle}
    \newcommand{\bra}[1]{\langle #1 |}
    
	\newcommand{\proj}[2]{\ket{#1}\bra{#2}}
	\newcommand{\pure}[1]{\proj{#1}{#1}}
 
	\newcommand{\set}[1]{\left\{ #1 \right\}}
	\newcommand{\abs}[1]{\left| #1 \right|}
	\newcommand{\tr}{\operatorname{Tr}  }
    
	\newcommand{\End}[1]{\operatorname{End}(#1)}

      \newcommand{\Set}[1]{\mathcal S(#1)}
      \newcommand{\sSet}[1]{\mathcal S_\leq (#1)}

	\newcommand{\id}{\mathbbm{1}}
	\newcommand{\hmin}{ H_{\min} }
	\newcommand{\hmax}{ H_{\max} }  
	
	\newcommand{\Hh}{H} %{H_H}
	\newcommand{\Dh}{D_H}

\newcommand*{\eps}{\varepsilon}
\newcommand*{\half}{\frac{1}{2}}

\newcommand*{\I}{\mathcal{I}}
\newcommand*{\E}{\mathcal{E}}

\newcommand{\cT}{\mathcal{T}}

\theoremstyle{plain}
\newtheorem{thm}{Theorem}
\newtheorem{lemma}[thm]{Lemma}
\newtheorem{corollary}[thm]{Corollary}

\theoremstyle{definition}
\newtheorem{defn}{Definition}
	\newcommand{\green}[1]{\textcolor{OliveGreen}{#1}}

	\newcommand{\ie}{i.e., }

	\newcommand{\cj}{Choi-Jamio\l{}kowski }
\newcommand{\flag}[1]{\green{ [#1]}}

	\newcommand{\eref}[1]{Eq.\ \ref{#1}}
	\newcommand{\fref}[1]{Fig.\ \ref{#1}}

\begin{document}

\title{Relative thermalization}

\author{L{\'i}dia \surname{del Rio}}
\affiliation{Institute for Theoretical Physics, ETH Zurich, Switzerland.}

\author{Adrian Hutter}
\affiliation{Department of Physics, University of Basel.}
\affiliation{Centre for Quantum Technologies, National University of Singapore, Singapore.}
\affiliation{Institute for Theoretical Physics, ETH Zurich, Switzerland.}

\author{Renato Renner}
\affiliation{Institute for Theoretical Physics, ETH Zurich, Switzerland.}

\author{Stephanie Wehner}
\affiliation{Centre for Quantum Technologies, National University of Singapore, Singapore.}

\begin{abstract} 
When studying thermalization of quantum systems, it is typical to ask whether a system interacting with an environment will evolve towards a local thermal state. 
Here, we show that a more general and relevant question is ``when does a system thermalize relative to a particular reference?"
By \emph{relative thermalization} we mean that, as well as being in a local thermal state,  the system is uncorrelated with the reference.  
We argue that this is necessary in order to apply standard statistical mechanics to the study of the interaction between a thermalized system and a reference. 
We then derive a condition for relative thermalization of quantum systems interacting with an arbitrary environment.
This condition has two components:
the first is state-independent, reflecting the structure of invariant subspaces, like energy shells, and the relative sizes of system and environment;
the second depends on the initial correlations between reference, system and environment, measured in terms of conditional entropies.
Intuitively, a small system interacting with a large environment is likely to thermalize relative to a reference, but only if, initially,  the reference was not  highly correlated with the system and environment. Our statement makes this intuition precise, and we show that in many natural settings this thermalization condition is approximately tight.
Established results on thermalization, which usually ignore the reference, follow as special cases of our statements.
\end{abstract}

\maketitle

\section{The case for relative thermalization}
\label{section:examples}

\subsection{Subjectivity in thermodynamics}

Thermodynamics was originally developed to study and improve the performance of steam engines: to turn the heat of a gas into work, as efficiently as possible. Today, it is also being applied to study heat and work flows in the micro and nano regimes. In fact, advances in the manipulation of small systems have allowed us to extract work from systems such as quantum dots and trapped ions \cite{Esposito2012a, Blickle2011}. 
Yet, thermodynamics as a science is still adapting to this new regime, and it still bears some of the traits of the gaseous systems for which it was first designed.                                                                                                                                                                                                                                                                                                                                                                                     
For example, the information available about the state of a gas used to be limited and objective: we would measure the temperature, pressure and volume of a gas, but we could not keep track of each individual particle. 
Crucially, all conceivable  observers had access to the same information about the state of the system, and could manipulate it in equivalent ways|like letting a gas expand to obtain work.
And yet, since very early on, several thought experiments have challenged the idea that thermodynamics should be objective. In 1871 James Maxwell realized that a ``demon'' able to measure the position and velocity of the particles of a gas could extract more work from it than the typical observer implicit in standard thermodynamics \cite{Maxwell1871}.
Picking up on Maxwell's idea on the power of information, Le\'o Szil\'ard imagined a partitioned box with a single-particle gas on one side. Depending on their information on the location of the particle, two observers would extract different amounts of work from the very same box~\cite{Szilard1929}.  

In spite of those examples, the idea that information about physical systems should be limited and objective became the core of a new discipline,  statistical mechanics.
Indeed, the fundamental postulate of statistical physics is the assumption that systems in contact with an environment equilibrate to a thermal state of maximum entropy, or ignorance. 
More precisely, the postulate states that, in equilibrium, an isolated  system is equally likely to be in any of the microstates that satisfy a given constraint, usually energy conservation. Under certain reasonable conditions, this probabilistic mixture of microstates results in the familiar Gibbs state.  It is implicit in this assumption that thermalization is independent of any external observer.

In recent years, there has been immense progress on the derivation of this  postulate  from first principles of quantum mechanics. Most studies to date focus on deriving conditions that lead to the local thermalization of a quantum system in contact with a large environment \cite{Popescu2006, Linden2009, Brandao2011, Masanes2013, Riera2012}. 
However, knowing if a system is locally thermalized is not enough for many practical applications. In what follows, we show that, even if a system is in a local thermal state, it will not necessarily act as a thermal bath towards all reference systems (or observers), and it is imperative to consider a stronger notion of thermalization. In the remainder of Section~\ref{section:examples}, we introduce the concept of thermalization of a system relative to an explicit reference and justify its relevance; in Section~\ref{section:results}, we study when relative thermalization is achieved.

\subsection{Defining relative thermalization}

\label{sec:setting}

Consider an arbitrary quantum system, $S$, which may be in contact with an environment $E$. Both $S$ and $E$ can be correlated with a reference  system $R$, and such correlations are described by the initial global state $\rho_{SER}$. 
In general, the system and its environment may be subject to physical constraints, like energy conservation. We represent an arbitrary constraint via a subspace $\Omega \subseteq S\otimes E$ of the joint Hilbert space of system and environment; for instance, $\Omega$ could be an energy shell. The time evolution of system and environment is given by a unitary operation in $\Omega$, $U_\Omega$.

\begin{defn}[Relative thermalization] \label{def:LocalTherm}
 Let $S$, $E$ and $R$ be quantum systems, and let $\Omega \subseteq S\otimes E$ be a subspace representing a physical constraint. The global system is in a state $\rho_{SER}$ of $\Omega \otimes R$.  
 We say that $S$ is \emph{thermalized relative to $R$} if $\rho_{SR}=\pi_S \otimes \rho_R$, where $\rho_R$ is arbitrary and $\pi_S$ is a local microcanonical state, defined as
\begin{align*}
 \pi_S := \tr_E \, \pi_\Omega,
\end{align*}
where  $\pi_\Omega := \frac{\id_\Omega}{|\Omega|}$ is the fully mixed state of $\Omega$.
More generally, we say that $S$ is \emph{$\delta$-thermalized relative to $R$} if it is $\delta$-close to the relative thermalized state, according to the trace distance,
\begin{align*}
 \frac12 \, \| \rho_{SR} - \pi_S \otimes \rho_R \|_1 \leq \delta .
\end{align*}
\end{defn}

Note that Definition~\ref{def:LocalTherm} does not require the global state of $S$ and $E$ to actually be  $\pi_\Omega$; only the reduced state of $S$ needs to be (approximately) microcanonical, and decoupled from the reference. 
Under certain natural conditions, like weak coupling, $\pi_S$ approximates a  Gibbs state \cite{Riera2012}.\footnote{We do not address the question of the exact form of $\pi_S$ here; we refer to it as a local thermal state independently of the notion of temperature. Note however that the constraint $\Omega$ determines the temperature of the thermal state.} 

In order to better understand this definition, we note that, as knowledge is relative, so is thermalization. An observer who can only measure a few parameters of the system might see it as thermalized, while someone with more precise measurement instruments (like Maxwell's demon) may see a well-defined microstate. The knowledge of different observers may be modelled by distinct reference systems. For example, we can think of the following state of  $SE$ and two references $R$ (the memory of the demon) and $R'$ (the memory of an observer that only measures enough parameters to determine $\Omega$),
\begin{align*}
 \rho_{SRR'} = \left( \sum_{i=1}^{\abs{\Omega}} \frac1{\abs \Omega} \tr_E{\pure i_\Omega} \otimes \pure i_R \right) \otimes \pure{0}_{R'}.
\end{align*}
Clearly $S$ is locally thermal, and it is also thermalized with respect to the reference $R'$: the reduced state of $SR'$ is precisely $\pi_S\otimes \pure{0}_{R'}$. However, $S$ is not thermalized with respect to $R$: the two are classically correlated.\footnote{
In a classically correlated state $\sum_i \sigma_i \otimes \pure i_R$, the reference $R$ can be seen as a classical memory, saving the value $i$ that tells us the state $\sigma_i$ of $S$. 
See also Section~\ref{sec:explicitref}.
} 
In what follows, we will show that this difference has actual physical consequences: it tells us whether a system acts as a heat bath towards a reference.

% xxxxxxxxxxxxxxxxxxxxxxxxxxx Examples xxxxxxxxxxxxxxxxxxxxxxxxxxxxxxxxxxxxxxxxx

\subsection{A first example: anomalous heat flow}

\begin{figure}[ht]
   \centering
   \includegraphics[width=6cm]{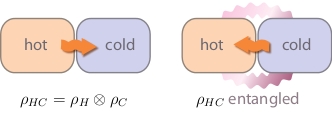}
   \caption{{\bf Anomalous heat flow.} If two thermal bodies are put in contact, heat normally flows from the hotter body to the colder one. However, it could be that the two systems are correlated, while still presenting local thermal states. If those correlations are strong enough (for instance if they are highly entangled), heat may flow from the colder to the hotter body. There is no contradiction with the second law, if one formulates it in terms of relative thermalization, because the two bodies are not thermal \emph{relative to each other}.}
   \label{fig:anomalous}
\end{figure}

Consider two systems $H$ and $C$, each in a local thermal state (in this case, their reduced density operators are Gibbs states of different temperatures, $\pi_H$ and $\pi_C$).
If we put the two systems in thermal contact, we would expect heat to flow from the hotter bath, $H$, to the colder one, $C$ (see \fref{fig:anomalous}). 
However, if $H$ and $C$  are highly entangled, one can observe an anomalous heat flow from $C$ to $H$ \cite{Partovi1989, Jennings2010, Jevtic2012}. 
The clue to understand this phenomenon is that $H$ and $C$ are not truly heat baths \emph{with respect to each other}. 
In our language, if we take $C$ to be the reference, it is clear that $H$ is not thermalized relative to it (and vice-versa),  because their joint state is not of product form, $\rho_{HC} \neq  \pi_H \otimes \pi_C$.   
Nevertheless, $H$ can still act as a normal heat bath towards a different reference system $R$, provided they are not initially correlated ($\rho_{HR} = \pi_H \otimes \rho_R$).

Clausius' formulation of the second law of thermodynamics states that heat cannot flow from cold to hot bodies~\cite{Clausius1867}. 
When this law was originally suggested, correlations between such systems were yet to be studied, and even today the law is implicitly interpreted as ``whenever two systems in local thermal states are put in contact, heat cannot flow from the colder system to the hotter one''. This reading, however, cannot be correct, given the violation brought about by anomalous heat flows.
In order to clarify its meaning, Clausius' law could be reformulated as ``whenever two systems which are \emph{thermal relative to each other} are put in contact, heat will not flow from the colder to the hotter body'' (up to fluctuations \cite{Jarzynski2011}).\footnote{The lyrics of Flanders and Swann's ``First and Second Law'' might prove trickier to adapt.}

\begin{figure*}[ht]
   \centering
   \includegraphics[width=5cm]{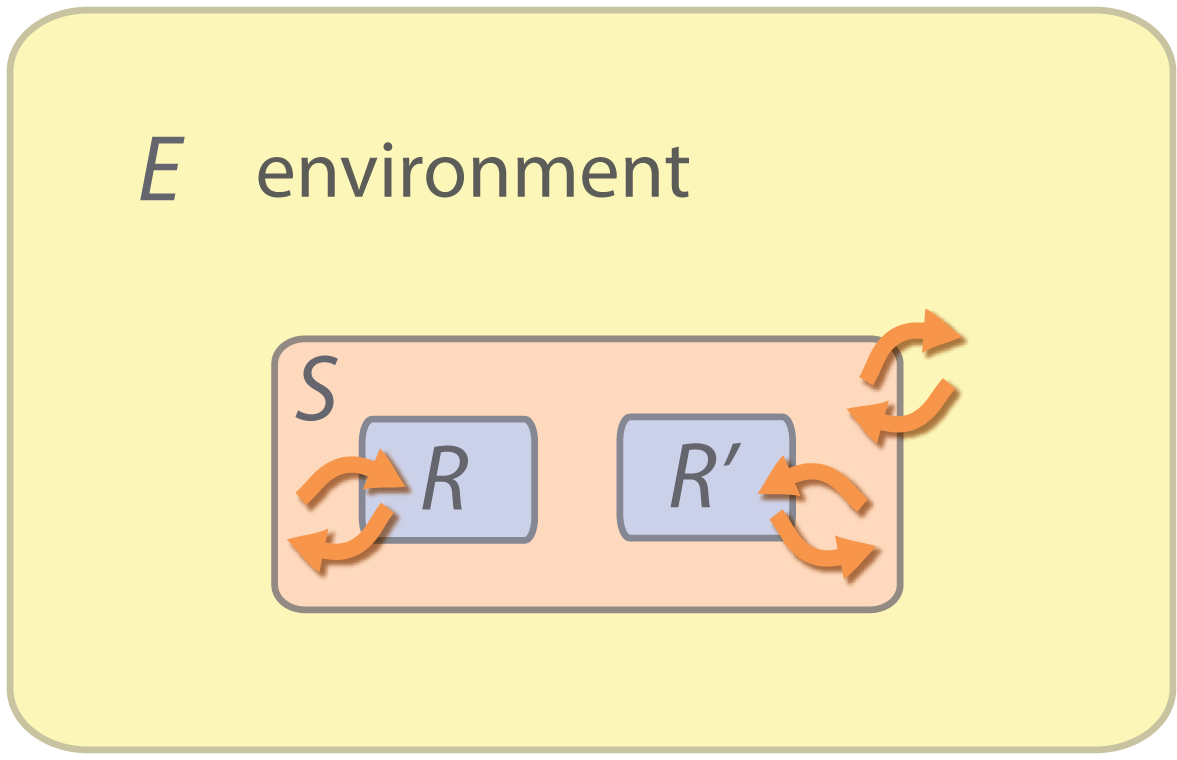}
   \caption{ {\bf Thermal noise.}
   To illustrate that local thermalization of a system $S$ is not enough to ensure that it will act as a source of white noise towards a reference device, consider the following toy example. Let $S$ be the part of an environment that is in contact with  two reference devices, $R$ and $R'$ (for instance, two quantum memories).
Suppose that the initial state of $SRR'$ is $\rho_{SRR'} =  \pure{\psi}_{SR} \otimes \rho_{R'}$, where $\ket\psi$ is  entangled between $S$ and $R$,
$\ket{\psi} = Z^{-1/2}  \sum_i \exp(- E_i/ 2 k T) \, \ket{i}_S \ket{i}_{R},$
with $Z=\sum_i \exp(- E_i/  k T)  $. 
The reduced state of $S$ is a local thermal state of temperature $T$, 
$\rho_S = Z^{-1} \sum_i \exp(- E_i/  k T) \,  \pure{i} = \pi_S . $
Now let us see if $S$ acts as a source of thermal noise if we let it interact with each of the two devices $R$ and $R'$. First  we look at $R'$. The joint state of $S$ and $R'$ is $\pi_S\otimes \rho_{R'}$, and we can say that $S$ is thermalized with respect to $R'$. If the temperature $T$ is high, $\pi_S$ is a very mixed state, which acts as a source of noise towards  $R'$: a joint unitary evolution of $S$ and $R'$ is likely to increase the entropy of $R'$. 
Now to device $R$. Since $S$ and $R$ are highly correlated, many joint evolutions of $S$ and $R$ will decrease the entropy of $R$|not the typical effect of a thermal bath. Indeed, in the limit $T \to \infty$, $\ket{\psi}_{SR}$ is a maximally entangled state, and no global evolution of $S$ and $R$ can increase the entanglement between the two, and therefore the entropy of $R$.  
   }
   \label{fig:noise_models}
\end{figure*}

\subsection{Second example: noise models}

In the field of quantum error correction we already think implicitly of relative thermalization. 
Imagine that you have a quantum device, like a small memory, and you would like to store quantum information on it. In general, this is a hard problem: the device is in contact with an environment, which interacts with the device, introducing errors. In order to build and analyse stable error-correction schemes, it is essential to model the interaction between the device and its environment. 

Most error-correction codes for quantum devices assume a noise model under which the environment acts as a source of randomness towards the device (like a heat bath of high temperature). Furthermore, it is often assumed that there are no memory effects: each new interaction between the device and its environment (that is, each new error) is independent of previous interactions~\cite{Gottesman1997}.  We call this a Markovian error model; an example would be a Markovian depolarizing channel.\footnote{This means that, for each time step, the interaction between device and environment is the same, and independent of previous interactions; we could also demand that, at a given time, errors induced in different parts of the device be independent and identically distributed (i.i.d.).} 
To get a physical intuition for this kind of model, we can think of an environment that consists of a thermal gas. If this environment is large enough, a gas particle that interacts with our device (and might become correlated with it) will probably get scattered away quickly, and will not interact with the device again. This way, the device is always interacting with fresh, uncorrelated particles from the gas.
Effectively, this means that each particle interacting with the device is in a local thermal state that is independent from the device|in other words, it is thermalized relative to the device.  

Using our notion of relative thermalization, we can reformulate the assumption of Markovity. 
Let $S$ be the subsystem of the environment that is in contact with the device (for instance, a thin layer of gas around the device, see \fref{fig:noise_models}).  
We require that $S$ be thermalized relative to the device before each interaction with the device. Markovity follows, and current error-correction schemes will work whenever this condition is satisfied.\footnote{
See \cite{Alicki2013, Jouzdani2013} for studies of how realistic environment models affect established fault-tolerance schemes, in particular what happens when the environment does not thermalize quickly enough with respect to the device, and the two remain correlated.}

\subsection{The need for an explicit reference in the quantum setting}
\label{sec:explicitref}

If a classical reference $R$ is correlated with a system $\Omega$, we can describe their joint state as a classical-quantum density matrix, $\rho_{R\Omega} = \sum_x p_x \ \pure x_R \otimes \rho_\Omega^x$. Crucially, this means that for each fixed value of classical knowledge $x$ in the reference, we can assign a reduced density matrix $\rho_\Omega^x$ to system $\Omega$. In other words, if an observer reads off the reference (which they can do without disturbing the global state) and finds it to be in state $\ket x$, then they know that $\Omega$ is in state $\rho_\Omega^x$|we may call it the state of $\Omega$ conditioned on knowledge $x$ in the reference. On a similar note, the von Neumann entropy of $\Omega$ conditioned on the reference $R$ is simply the average of the entropies of the conditional states, $H(\Omega|R)_\rho = \sum_x p_x \ H(\Omega)_{\rho^x}$. In particular,
 this entropy is always non-negative, because the non-conditional entropy $H(\Omega)_{\rho^x}$ is never negative.

This way of thinking about the knowledge stored in a reference breaks down in the quantum world. Imagine now that both the reference and the system are quantum-mechanical; in particular, they could be entangled, for instance in state $\rho_{R\Omega} = \pure \Psi$, with $\ket\Psi_{R\Omega} = \sum_x \sqrt{p_x} \ \ket x_R \otimes \ket x_\Omega$. In this case,  we cannot define a ``conditional state'' of $\Omega$ for each fixed value of knowledge in $R$ (in fact such fixed values do not exist).
A simple way to see this is by looking at the conditional entropy $H(\Omega|R)_\rho$: if we could write it as an average  $\sum_\alpha p_\alpha \ H(\Omega)_{\rho^\alpha}$, then it would be positive, but the entropy of entangled states like $\ket\Psi_{R\Omega}$ is actually negative. 

In the setting of thermalization, one may argue that, if we have a classical reference $R$, we may simply read its state $\ket x$, consider the conditional state $\rho_\Omega^x$ and study local thermalization of a subsystem $S$ starting from that state. Clearly, if, after an evolution of $\Omega$, the final state of $S$ is thermalized, it is also decoupled from the reference. In fact, this is implicitly done in the current literature, when we talk about the ``initial knowledge'' of the state of $\Omega$. 
However, we cannot take this approach when the reference is itself a quantum system, a more general and natural setting than imposing classicality on the reference|in the examples we saw, $R$ was simply another system that was, at some point, in contact with $\Omega$, and became correlated with it. In order to study the evolution of $\Omega$ with respect to $R$ in this general framework, we need to consider their joint density matrix.

\section{Technical results: when is relative thermalization achieved?}
\label{section:results}

\subsection{Summary and related work}

\begin{figure}[ht]
   \centering
   \includegraphics[width=7cm]{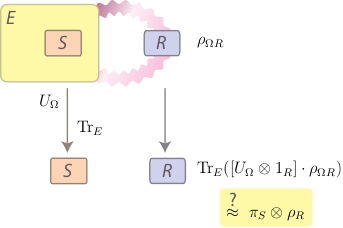}
   \caption{{\bf Setting.} The initial state $\rho_{\Omega R}$ evolves as $(U_\Omega \otimes \id_R ) \ \rho_{\Omega R} \ (U_\Omega^\dagger \otimes \id_R ) $, where $U_\Omega$ is a  unitary acting on $\Omega$.
   Now we consider only the reduced state of the subsystem $S$ and the reference $R$, and compute its distance to the decoupled thermal state $\pi_S \otimes \rho_R$. If this distance is small, then the final state on $S$ is approximately thermalized relative to $R$ (see Definition~\ref{def:LocalTherm}).
   }
   \label{fig:diagram}
\end{figure}

We consider the general setting described in \fref{fig:diagram}, where a system $S\otimes E$ is subject to a physical constraint $\Omega$. 
An observer (or reference) $R$ may hold quantum information about $S\otimes E$: this is expressed in the initial state $\rho_{\Omega R}$. We want to know what kind of initial states and unitary evolutions in $\Omega$ lead to thermalization of the subsystem $S$ relative to $R$ (according to Definition~\ref{def:LocalTherm}).

Our approach generalizes recent efforts to study local  equilibration of quantum systems  \cite{Popescu2006, Linden2009, Brandao2011, Masanes2013, Riera2012}. 
These studies have shown that, even if the global system $S\otimes E$ is not thermal, the reduced state of $S$ may equilibrate to $\pi_S$.
They prove that the relative size of the system $S$ compared to the environment $E$ affects thermalization: small systems in a large environment almost always equilibrate to the microcanonical state. Another factor that determines how quickly $S$ thermalizes is the structure of the Hamiltonian of $SE$: the systems must be fully interacting and it helps if their joint evolution drives them through many different states. There have also been converse results, on states that do not equilibrate~\cite{Gogolin2011a, Hutter2013}.
The results of~\cite{Popescu2006, Linden2009, Brandao2011, Masanes2013, Riera2012}, originally derived through measure concentration techniques and from properties of the system's Hamiltonian, emerge here as direct consequence of our general approach, in the special case where the reference $R$ is  classical. 

We show that, apart from the physical conditions for thermalization found in the literature, there is another fundamental factor for relative thermalization, namely the initial correlations between the reference and the system and its environment. An observer that knows little about the initial state of $S \otimes E$ will see $S$ thermalize, even if $S$ is not much smaller than $E$; on the other hand, in the extreme case where the reference is highly entangled with $SE$, even a small subsystem $S$ will not appear to thermalize.

\begin{figure*}[ht]
   \centering
   \includegraphics[width=6cm]{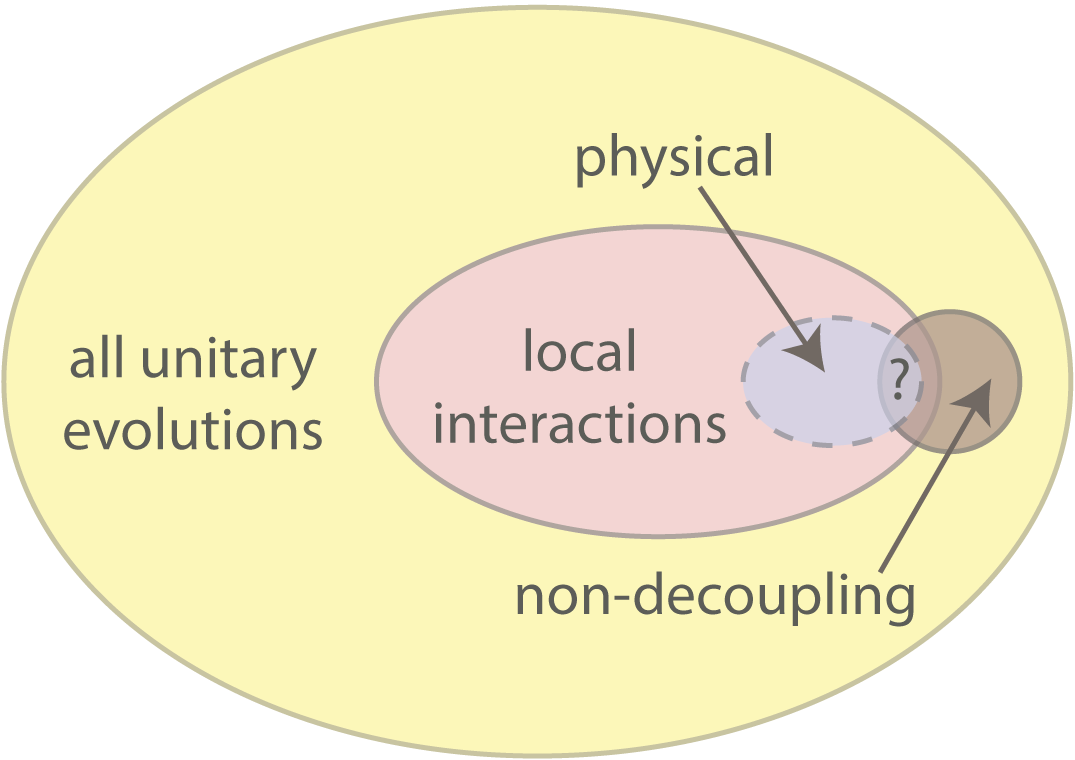}
   \caption{ {\bf Typicality of relative thermalization.} 
   Physically, only one unitary $U_\Omega$ is realized as we let our systems evolve for a certain period of time. 
   In Theorem~\ref{thm:typical_main} we state that most unitaries, according to the Haar measure, lead to relative thermalization. This means that, if all we know about $U_\Omega$ is that it is a unitary in $\Omega$, it is highly likely, from our point of view, that $U_\Omega$ will  thermalize $S$ relative to $R$. 
   Usually, though, we know more about $U_\Omega$, for instance, that it is induced by a given local Hamiltonian.
   As the set of all unitaries in $\Omega$ is full of  operators that are unrelated to our physical setting (like non-local evolutions, ruled out by our knowledge), it is desirable to obtain similar probabilistic statements about smaller sets that still contain $U_\Omega$ (like those generated by local interactions).
   This is possible, because the decoupling approach~\cite{Hayden2007, Dupuis2010, Dupuis2010a, Szehr2011} used to obtain our results is very general, and can be applied to more physical sets of unitaries, consisting of local two-body interactions \cite{Brandao2012,Brown2012,Brown2013}, or time-independent Hamiltonians \cite{Masanes2013, Brandao2011, Brandao2012}. 
   For a more detailed discussion, see Section~\ref{section:local}. 
   }
   \label{fig:classes_unitary}
\end{figure*}

We derive typicality statements, of the form ``if such entropic condition stands, then most evolutions in $\Omega$ lead to thermalization of $S$ relative to $R$'' (Theorem~\ref{thm:typical_main}). 
In the usual thermodynamic limit of a small subsystem $S$, this result is tight, in the sense that, if a similar entropic relation is not satisfied, then no unitary evolution in $\Omega$ can lead to relative thermalization (Theorem~\ref{thm:converse_main}). See \fref{fig:classes_unitary} for a discussion on the role of typicality in our results.

\subsection{Entropy: measuring correlations}
\label{sec:entropy}

Our results rely on decoupling~\cite{Hayden2007, Dupuis2010, Dupuis2010a}, which is tightly characterized by smooth entropies, a natural class of entropies quantifying correlations between quantum systems in single-shot settings (see Appendix~\ref{appendix:entropy}). 
From this class, we choose a particular conditional entropy,
 $\Hh^\eps(\Omega|R)_\rho$, to express our results \cite{Dupuis2012}.
 For the sake of space, we define and characterize $\Hh^\eps$, sometimes called the hypothesis-testing entropy, in Appendix~\ref{appendix:entropy}. 
 For now, keep in mind that conditional entropies measure our uncertainty about the exact state of $\Omega$, given access to system $R$, for the quantum state $\rho_{\Omega R}$.  
 The parameter $\eps \in [0,1]$ is related to a small probability of error, or, in other words, to our willingness to ignore highly unlikely events, like the possibility of a shattered glass coming together again in a split second. In many natural scenarios, we want $\eps$ to be small but non-zero.\footnote{A note on the operational meaning of  $\Hh^\eps$ within information theory. For small $\eps$, $\Hh^\eps(\Omega|R)_\rho$ can be used to quantify the amount of pure randomness that can be extracted from the state in $\Omega$, such that it is independent from $R$. 
For large $\eps$,  it is related to the task of data compression (or erasure of information) in the presence of a quantum memory $R$. For the experts, $\Hh^\eps$ approximately interpolates between the smooth min- and max-entropies (see Appendix~\ref{appendix:entropy} and \cite{Dupuis2012}).}

To give an idea of the values that this entropy takes, consider the limit $\eps \to 0$. Then, $\Hh^\eps (\Omega|R)_\rho$ is zero if $\rho_\Omega$ is pure, is at most $\log_2 |\Omega|$, which is achieved for the fully mixed, decoupled state  $\rho_{\Omega_R} = \frac{\id_\Omega}{|\Omega|} \otimes \rho_R$, and becomes negative  if the $\Omega$ and $R$ are entangled, with a minimum at $-\log_2 |\Omega|$ for maximally entangled states.
 $\Hh^\eps(\Omega|R)_\rho$ has the natural properties expected from conditional entropy measures, like the data-processing inequality, which states that locally processing information in $R$ cannot give us more knowledge about $\Omega$.
 
$\Hh^\eps$ converges to the familiar von Neumann entropy in the asymptotic limit of many independent copies of the global system $\Omega \otimes R$,
$$ \lim_{n \to \infty} 
		\frac{1}{n} \Hh^{\eps}(\Omega^{\otimes n} | R^{\otimes n} )_{\rho^{\otimes n}}
			= H(\Omega | R )_{\rho} .$$
In information theory, this limit is applied to many sequential uses of the same resources, or repetitions of an experiment | which is why the von Neumann entropy is used to characterize the success rate of information-processing tasks.
In thermodynamics, we do not always have the luxury of arbitrarily repeating experiments (like letting
a cup of coffee thermalize several times), and are usually interested in predictions for a single instance of an event (what is the probability that this cup of coffee cools down now?). 
The same limit emerges, however, in the treatment of large systems made out of many uncorrelated subsystems, like an ideal gas.

\subsection{Achievability of relative thermalization}

Theorem~\ref{thm:typical_main} gives us tight conditions to find a subsystem of $\Omega \in S\otimes E$ to be thermalized with respect to the reference after most unitary evolutions $U_\Omega$ (see \fref{fig:diagram}).  
It tells us that, under certain entropic conditions, only an exponentially small fraction of evolutions in $\Omega$ do not lead to relative thermalization.
In the theorem, $\abs \Omega$ stands for the dimension of Hilbert space $\Omega$, and $\delta$-relative thermalization refers to Definition~\ref{def:LocalTherm}.
A technical version of this statement can be found in Appendix~\ref{appendix:proof_direct}.

\begin{thm}[Thermalization of typical subsystems]\label{thm:typical_main}
Let $\rho_{\Omega R}$ be a quantum state in $\Omega\otimes R$, with $\Omega \subseteq S\otimes E $,
and let $\pi_\Omega= \frac{\id_\Omega}{\abs \Omega}$. 
Let $\eps, \delta >0 $.  
If the entropic relation
\begin{align}
 \Hh^{ 9   \eps } (SE|R)_\rho 
 >  \Hh^{1- \eps}(S)_\pi
 - \Hh^{\eps}(E)_\pi  
 + \mathcal{O}\left(\log \frac1{\eps + \delta} \right) 
 \label{eq:condition_main}
\end{align}
 holds, then, after a unitary evolution $U_\Omega$  of $\rho$ in $\Omega$, $S$ will be $\delta$-thermalized relative to $R$, 
 except for a fraction  $2\ e^{- \frac{ \abs{\Omega}}{ 16}  \delta^2}$ of the unitaries acting on $\Omega$, according to the Haar measure. 
\end{thm}

Note that the entropic terms on the right-hand side of (\ref{eq:condition_main}) are evaluated on the reduced states of the canonical state $\pi_\Omega$ | they depend only on the structure of the physical constraint $\Omega$, which is determined by factors like the Hamiltonian of $S\otimes E$. Therefore, we may bound these measures with state-independent quantities, such as the dimensions of $S$ and $\Omega$ (see \eref{eq:dimensions}).  On the left-hand side, we have $\Hh^{9\eps} (SE|R)_{\rho}$, which depends on the global initial state. This term gives us an information-theoretical condition for relative thermalization: if the reference is not highly correlated with $S \otimes E$, then a typical evolution in $\Omega$ is likely to ``sweep'' correlations with $S$ to the environment, leaving $S$ thermalized relative to $R$.

Since we are usually interested in the limit of small $\eps$, it might at first appear concerning that our bounds (the right-hand side of the condition from Theorem~\ref{thm:typical_main}) diverge in that limit. 
However, the divergence is only logarithmic in $\eps$, and does not depend on the size of the systems involved. 
The entropic terms, on the other hand, grow  with the size of the systems. In the thermodynamic limit of large systems, the logarithmic divergence is negligible. 

\begin{figure*}[ht]
   \centering
   \includegraphics[width=9cm]{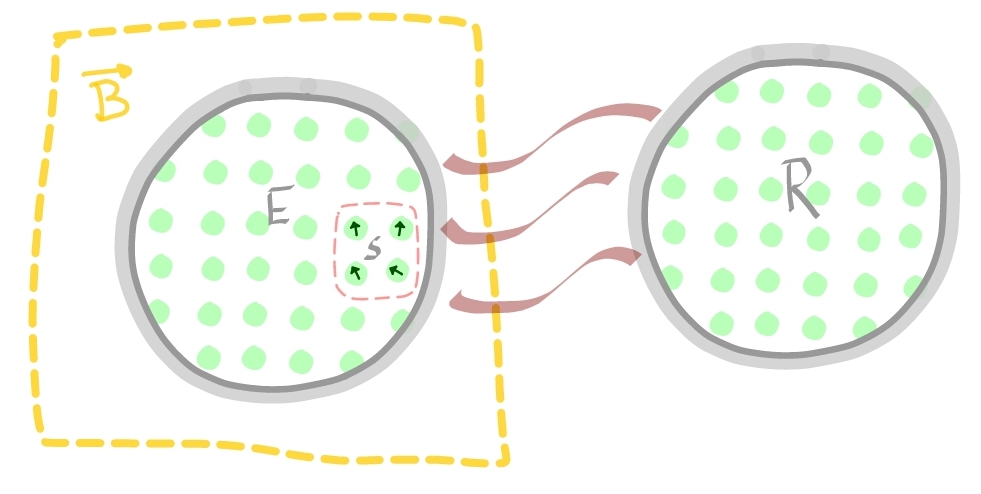}
   \caption[LoF entry] {
   {\bf Application of our results.} 
   Consider a system of $N$ weakly interacting spins, subject to the Hamiltonian $\hat H = \hat H_0 + \hat V$, where $\hat H_0=J \sum_i \ket{ \! \! \uparrow}\bra{\uparrow \! \!}_i$ and  $\hat V$ is a random nearest-neighbour perturbation that conserves the total spin (with $|\hat V| \ll |\hat H_0|$); this system is also studied in the preprint version of \cite{Popescu2006}.
   We select $\alpha \, N$ of those spins to be our subsystem $S$, while the remaining $(1-\alpha) N$ spins are called the environment $E$. 
   In addition, the spins of $S \otimes E$ may be correlated with a reference spin system $R$. 
   
   We want to study thermalization of $S$ relative to $R$, for an arbitrary initial state $\rho_{SER}$.
   Note that the energy subspaces of $S\otimes E$ are invariant under time evolution ruled by $\hat H$; therefore we will look at states that lie in one of these invariant subspaces. For mixtures and superpositions over different subspaces, the results follow by linearity. Each energy shell $\{\Omega_k\}$ is generated by states with a fixed number $k$ of spins up,  $\Omega_k= \mbox{span} \{\ket\Psi_{SE}:\  \hat H_0 \, \ket\Psi = k \, J\, \ket\Psi\}$. 
   We apply the condition for relative thermalization given by (\ref{eq:dimensions}) to states  $\rho_{SER}$  with $\rho_{SE} \in \End{\Omega_k}$ for some $k$.  The dimension of $S$ is $2^{\alpha  N}$, while $|\Omega_k| = \binom N {k}$. For large $N$, $\log \binom N {k} \approx N \, h( k/ N)$, where $h(p) = - p \, \log p - (1-p) \log(1-p)$ is the binary entropy of $p$. 
   Plugging these dimensions into (\ref{eq:dimensions}), we obtain the condition: 
   if
   $ H^\eps (SE|R)_\rho > N [2 \alpha - H( k /N)] $, 
   then $S$ will be $\delta$-thermalized relative to $R$ after most evolutions. In particular, if $\hat V$ is generic enough, our system can be modeled as a typical local circuit, such that relative thermalization holds for most of the time, after an initial equilibration period (see Section~\ref{section:local}). 
   Conversely, if $\Hh^\eps (SE|R)_\rho  < N (1-\alpha)$, then no evolution in $\Omega_k$ leads to relative thermalization of $S$ (Theorem~\ref{thm:converse_main}). As the systems in this example are large, we neglect the size-independent logarithmic terms in $\eps$ and $\delta$.
   }
   \label{fig:example}
\end{figure*}

To give an idea of the dimension of the entropic terms, we can find a weaker condition (see Appendix~\ref{appendix:dimensions}). At least the same fraction of unitary evolutions as in Theorem~\ref{thm:typical_main} leads to relative thermalization, as long as 
\begin{align}
 \Hh^{\eps} (SE|R)_\rho  
 >  \log  \frac{|S|^2}{|\Omega|}  
 +\mathcal{O}\left(\log \frac1{\eps + \delta} \right).
 \label{eq:dimensions}
\end{align}
See \fref{fig:example} for a simple example.

\subsection{Converse}
\label{sec:converse}

Sometimes, the reference is so correlated with $S\otimes E$ that no evolution in $\Omega$ can decouple $S$ from it. 
Theorem~\ref{thm:converse_main} characterizes the states that can never achieve relative thermalization. The setting is the same as in Theorem~\ref{thm:typical_main} and \fref{fig:diagram}.
\begin{thm}
  \label{thm:converse_main}
  Let $\rho_{\Omega R}$ be a quantum state in $\Omega\otimes R$, with $\Omega \subseteq S\otimes E $,
  and let $\pi_\Omega= \frac{\id_\Omega}{\abs \Omega}$.
  Let $\delta, \eps > 0$.
  If the entropic condition
  \begin{align}
   \Hh^{\eps  }(S E |R)_\rho 
  < - \Hh^1(E)_\pi + \mathcal{O}\left(\log \frac1{\eps + \delta} \right) 
  \label{eq:condition_converse}
 \end{align}
 holds, then no unitary evolution of $\rho$ in $\Omega$ can leave $S$ $\delta$-thermalized with respect to $R$.
\end{thm}

Note that (\ref{eq:condition_converse}) is close to a converse of the direct bound (\ref{eq:condition_main}), in the typical case of a large environment $E$ and small subsystem $S$, when it is reasonable to neglect a term of the order $\log \abs S$.  
In other words, the conditions for relative thermalization are tight in this typical setting. 
Observe however that, in order to achieve the converse bound in the typical setting of small system and large environment, the reference $R$ must be highly entangled with $S \otimes E$. In this case, the entropy $\Hh^\eps(SE|R)_\rho$ becomes negative and may cancel out $\Hh^1(E)_\pi$ | a simple example is a demon with maximal quantum knowledge about the initial state of $\Omega$.\footnote{In particular, the reference $R$ must be approximately as large as the environment itself, since $\Hh^\eps(SE|R)_\rho \geq - \min \set{\log |R|, \log |\Omega|}$.}  Intuitively, if the reference is very entangled with $S$ and $E$, then there is no unitary evolution that can move all of the entanglement to the environment. 
A technical version of this statement can be found in Appendix~\ref{appendix:converse}; there, we see that, for most unitaries  $U_\Omega$, the bound of Theorem~\ref{thm:typical_main} is tight even when the subsystem $S$ is large.

\section{Generalizations and applications}
\label{section:discussion}

\subsection{Typical local interactions}
\label{section:local}

So far we have used the
Haar measure to define the fraction of unitaries for which decoupling occurs. While this seems a natural choice from an information-theoretical perspective, it is desirable to find statements like Theorem~\ref{thm:typical_main} that apply to smaller sets of unitaries, ideally those containing only evolutions related to our physical problem (see \fref{fig:classes_unitary}).\footnote{In a celebratory analogy, imagine that you want to know whether a cryptic Vietnamese dish you were served is vegetarian. The statement ``only $5\%$ of Vietnamese dishes are vegetarian'' is more useful to you than the more generic information ``$30\%$ of all dishes cooked in the world are vegetarian''. Evidently, the world stands for the Haar measure, Vietnam for local interactions, your dish for the unitary $U_\Omega$ that was actually realized, and vegetarian for non-thermalizing. All statistics are wild guesses.
% Congratulations Marco! :)
}

A possible direction in the search for sets of more physical evolutions is given by local circuits. These can simulate, for instance, a chain of atoms in which, at each time step, every two neighbouring atoms undergo a joint unitary evolution, or a particle gas, where every two particles may interact locally at some point. These local quantum circuits were shown to achieve decoupling after an initial equilibration period | that is, after that period, a subsystem of a typical local circuit will thermalize with respect to a reference with high probability.\footnote{More precisely, rigid circuits like atom chains approximate $k$-designs~\cite{Brandao2012}, defined as a set of unitaries that reproduce the first $k$ moments of the Haar distribution~\cite{ Low2010, Brandao2012}. It turns out that decoupling results also apply to approximate $k$-designs~\cite{Hayden2007,Szehr2011, Brandao2012}. For non-rigid circuits like the particle gas, decoupling was proven directly \cite{Brown2013}.}
In other words, our results on relative thermalization apply to physical systems that can be described by two-body local interactions (up to logarithmic terms that do not scale with system size).

\subsection{Time scales}

The concept of thermalization relative to a reference may be applied to other aspects of thermalization. 
For instance, one may take a concrete time-independent Hamiltonian for our physical system, and look for the time scales of subsystem thermalization under the usual Schr\"odinger evolution.  Recent results in this field seem to indicate that the computational complexity of the Hamiltonian (that is, how easy it is to diagonalize it), is correlated with the time needed to achieve thermalization (see for instance \cite{Masanes2013, Riera2012}). This happens because complex Hamiltonians are characteristic of highly interacting, perturbed systems | the kind of places where subsystems quickly become entangled, and therefore locally mixed. 

Another possible angle is the study of subsystem thermalization in physical systems with an effective light cone (like spin lattices in which perturbations take some time to propagate). It was shown that the time scales of thermalization depend again on the size of the subsystem, as well as  on the emergent speed of light
\cite{Hutter2013}. 

In general, the study of time scales for thermalization can be extended to our setting, where there is an external quantum reference correlated with the system evolving. The question then is how long it takes for the reference to lose all information about the state of a given subsystem. One option is to apply existing decoupling techniques to this problem, for instance in the case of local circuits, where evolution time can be measured in terms of circuit size~\cite{Szehr2011, Brandao2012,Brown2012,Brown2013}). Another direction is to adapt existing techniques used in non-equilibrium thermodynamics to our setting.

\subsection{Thermalization under observables}

A different approach to thermalization is to ask whether isolated systems appear to thermalize under measurements \cite{Ududec2013, Reimann2012}. The setting: we have a system $\Omega$ in initial state $\rho_\Omega$, which undergoes a unitary evolution $U_\Omega$. Then we perform a measurement on $\Omega$, described by a positive-operator valued measure  $\set{M_x}$. The question is whether we can distinguish the actual state from the thermal state  $\pi_\Omega$, given only the  measurement statistics, \ie if $\tr(M_x \, [U_\Omega \cdot \rho_\Omega] ) \approx \tr (M_x \, \pi_\Omega)$ for all outcomes $x$. 

It is clear that if we could perform full state tomography then we could distinguish the two states, in particular if the initial state is close to pure. But often $\Omega$ is a large system and tomography is unpractical. For this question to have some operational meaning, we should restrict ourselves to measurements that can be implemented efficiently in the lab, for instance local or coarse-grained measurements. It turns out that under certain constraints on the complexity of the measurements allowed, most states will appear to thermalize after an initial equilibration period~\cite{Ududec2013}. 

This idea may be generalized to our setting, where we have side information about the initial state of $\Omega$ | our reference can be a quantum memory correlated with $\Omega$. The relevant question is whether this side information can help us distinguish the evolved state  of the system from a thermal state under feasible measurements. 

In order to reach a quantitative theorem, it is possible to directly apply the decoupling approach described here. More technically, when applying the decoupling theorem (see Appendix~\ref{appendix:decoupling}), one must choose the decoupling map that represents the measurement, instead of a partial trace over the environment.

\section{Conclusions}

Traditionally, thermodynamics deals with large-scale objects, and as a consequence, quantum correlations between systems can be neglected. This is because most degrees of freedom are irrelevant for the macroscopic behaviour of a system, or the performance of a heat engine: we are only interested in the average energy of a gas, or the position of a piston, and correlations are typically encoded in finer details of the particles' wave functions. 
However, as modern technologies  miniaturize to the nanoscale, a comprehensive understanding of the thermodynamics of small quantum systems is essential to identify and harness their power.
As the number of degrees of freedom decreases, correlations become more likely to influence the relevant parameters of an experiment, and can no longer be neglected. 
For example, correlations between heat baths have been shown to affect the performance of three-qubit heat engines~\cite{Brunner2013}. 
These engines only behave like traditional Carnot machines if the baths involved are thermalized relative to each other. 

Relative thermalization was also found to be crucial to prove Landauer's principle, which quantifies the work cost of information-processing tasks in physical systems~\cite{DelRio2011, Faist2012, Reeb2013}. In order to achieve Landauer's bound, it is necessary that the system of interest be decoupled from a thermal bath | otherwise we could exploit correlations with this ``bath'' to extract extra work. 

Our framework provides a wealth of open tasks for both the information-processing and the quantum thermodynamics communities, not least the generalization of known techniques to study different aspects of relative thermalization.

\begin{acknowledgments}
We acknowledge support from the Swiss National Science Foundation (LdR and RR, grant no.\ 200021-119868 and the NCCR QSIT), 
the Portuguese Funda\c{c}\~ao para a Ci\^{e}ncia e Tecnologia (LdR, grant no.\ SFRH/BD/43263/2008), 
the European Research Council (RR, grant no.\ 258932), and 
the National Research Foundation and the Ministry of Education, Singapore (AH and SW). 
We further acknowledge the Singapore programme Mathematical Horizons for Quantum Physics 2, and the COST Action MP1209 ``Thermodynamics in the quantum regime'' % lol
 for supporting our collaboration. 
We thank Jens Eisert, Philippe Faist, Christian Gogolin, Philipp Kammerlander  and Yeong-Cherng Liang for discussions and helpful comments on the manuscript.
Lastly, L{\'i}dia would like to thank Lea Kr{\"a}mer for lemmas over brunch, and Aleksey Fomins for exorcising epsilons, deltas and logarithms.
\end{acknowledgments}

% ............... Appendix ..............

\onecolumngrid
\newpage
\appendix

\begin{center}
 \Large{\sc{Appendices}}
\end{center}

\noindent{\bf A word on notation.} 
We use $\Set{A}$ to denote the set of density matrices acting on Hilbert space $A$, i.e.,
\begin{align*}
 \Set{A} = \set{ \rho \in \End{A}: \rho \geq 0, \tr \rho = 1 },
\end{align*}
where $\End{A}$ denotes endomorphisms on $A$.
Similarly, the set of subnormalized positive semi-definite operators ($\rho \geq 0$, $\tr \rho \leq 1$) is denoted by $\sSet{A}$.
For instance,  $\rho_{A B} \in \Set{A\otimes B}$ is the (possibly mixed) state of a bipartite quantum system, consisting of subsystems $A$ and $B$.  

The identity operator on Hilbert space $A$ is denoted by $\id_A \in \End{A}$, while the identity map acting on operators of $A$ is denoted by $\mathcal{I}_A \in \End{\End{A}}$. 

For simplicity, we use $U_A \cdot \rho_{AB}$ to denote $[U_A \otimes \id_B] \  \rho_{AB} \ [U_A^\dag \otimes \id_B]$.

We use $\log$ to denote the logarithm of base 2.

\section{Smooth entropy measures}
\label{appendix:entropy}

\subsection{Smooth min- and max-entropies}

Most of our technical proofs use conditional smooth min- and max-entropies \cite{Tomamichel2012, Renes2012, Konig2009, Datta2009, Renner2005}. These have convenient properties, used to derive the final form of our results (for example, duality, see \ref{eq:duality}). 
For a comprehensive discussion of these entropies, their properties and applications, we refer to \cite{Tomamichel2012}.

\subsubsection{Purified distance}

The purified distance~\cite{Tomamichel2010} is used to smooth the min- and max-entropies, and is defined for subnormalized states $\rho, \sigma \in \sSet{A}$.
Let us first recall the definition of fidelity,
\begin{align} \label{eq:fidelity}
 F(\rho, \sigma) := \| \sqrt \rho \, \sqrt \sigma \|_1,
\end{align}
where $\| \cdot \|_1$ is the $L_1$-norm.
The generalized fidelity is defined for subnormalized states as 
\begin{align}
 \bar{F}(\rho, \sigma) := F(\rho, \sigma) + \sqrt{(1-\tr \rho) (1-\tr \sigma)}.
\end{align}
Note that if at least one of the two states is normalized, we recover the usual fidelity.
Finally, the purified distance is defined in terms of the generalized fidelity, 
\begin{align}
 \label{eq:purified_distance}
 d(\rho, \sigma) := \sqrt{1 - \bar{F}(\rho, \sigma)^2}.
\end{align}
The purified distance is a metric, is invariant under purifications and extensions, and can only decrease under physical operations and projections  \cite{Tomamichel2010}.
It relates to the  trace distance as \cite{Tomamichel2010}
\begin{align}\label{eq:purified_vs_trace_distance}
  \half  \|\rho - \sigma \|_1  + \half |\tr\rho - \tr \sigma| \leq 	d(\rho, \sigma) \leq \sqrt{ \|\rho - \sigma \|_1 +|\tr\rho - \tr \sigma| }.
\end{align}

The $\eps$-ball around a positive operator $\rho \in \sSet{A}$ is defined as usually,
\begin{align*}
 \mathcal{B}^\eps (\rho) := \set{ \tilde \rho \in \sSet{A}:  d(\rho, \sigma)  \leq \eps}.
\end{align*}

\subsubsection{Smooth min-entropy}

The conditional smooth min-entropy $\hmin^\eps(A|B)_\rho$ can be used to quantify the size of a subsystem of $A$ that can be decoupled from $B$~\cite{Dupuis2010a}. 
In classical cryptography, it is applied to privacy amplification, giving us the length of a secret key that can be securely extracted from $A$ such that it is inaccessible to an adversary that controls system $B$. 
The non-smooth version of the min-entropy is defined as
\begin{align}
  \hmin(A|B)_\rho := 
  \sup_{\sigma_B \in \Set{B}}\
  \sup_{\lambda \in \mathbb{R}} \set{ \lambda:  2^{- \lambda} \id_A \otimes \sigma_B \geq \rho_{AB} }.
\end{align} 
In the particular case where the two systems are independent, $\rho_{AB} = \rho_A \otimes \rho_B$, the min-entropy is simply $-\log_2 \| \rho_A\|_\infty$, where  $\| \rho_A\|_\infty$ is the maximum eigenvalue of $\rho_A$.

Smoothing is made by optimizing the min-entropy over a small neighbourhood of $\rho$, according to the purified distance,
\begin{align}
  \hmin^\eps(A|B)_\rho := 
  \sup_{\tilde{\rho} \in \mathcal{B}^\eps (\rho)} 
    \hmin(A|B)_{\tilde{\rho} } .
\end{align} 
The \emph{smoothness parameter} $\eps \geq 0$ is usually chosen to be small but nonzero.
In most contexts, it corresponds to a small error probability.

\subsubsection{Smooth max-entropy}

The smooth conditional max-entropy $\hmax^\eps(A|B)_\rho$ can be used to quantify the number of bits necessary to reconstruct the state of system $A$, given quantum side information  $B$. In thermodynamics, it characterizes the work cost of erasure of $A$, given access to $B$~\cite{DelRio2011}. In classical information theory, the non-conditional max-entropy quantifies the compression rate of a random source $A$.
The non-smooth conditional max-entropy can be defined as
\begin{align}
	\hmax (A| B)_{\rho} 
	:= \sup_{\sigma_B \in \Set{B}} 
		\log_2 F\left(\rho_{A B}, \id_{A} \otimes \sigma_B\right)^2 , 
\end{align}
where $F$ is the fidelity (\eref{eq:fidelity}).
We smooth the max-entropy as we did with the min-entropy,
\begin{align}
  \hmax^\eps(A|B)_\rho := 
  \inf_{\tilde{\rho} \in \mathcal{B}^\eps (\rho)} 
    \hmax(A|B)_{\tilde{\rho} } .
\end{align}

\subsection{Generalized smooth entropy}

Our final results are expressed in terms of a generalized smooth entropy, introduced in \cite{Dupuis2012}. 
For $\eps > 0$, it is defined as 
\begin{align}
 \Hh^\eps(A|B)_\rho := -\Dh^\eps(\rho_{AB}||\id_A \otimes \rho_B),
 \label{eq:Heps_def}
\end{align}
where $\Dh^\eps$ is the hypothesis-testing relative entropy, defined as
\begin{align}
 2^{- \Dh^\eps(\rho||\sigma) } 
 :=\frac{1}{\eps}\inf_Q\set{\tr (Q \sigma) : 
  0\leq Q\leq \id \land \tr ( Q \rho) \geq \eps}.
  \label{eq:Deps_def}
\end{align}
This corresponds precisely to the setting of hypothesis testing: we are given one of two states $\rho$ and $\sigma$ at random, and we want to distinguish them with a single measurement, trying to be right on $\rho$ with probability at least $\eps$. 
We start from the set of all POVMs with two outcomes, $\set{Q, \id-Q}$: our guessing strategy is to say that the state is $\rho$ if we obtain $Q$ and $\sigma$ if we obtain $\id- Q$. 
First we restrict the set to those POVMs such that the probability of guessing correctly $\rho$ if the outcome is $Q$ is at least $\eps$.
To further optimize our overall guessing probability, we pick the POVM that minimizes the probability of obtaining $Q$ when measuring $\sigma$.

Further operational interpretations of the generalized smooth entropy come from its relation to the smooth min- and max-entropies, given below. 
In short, for small $\eps$ it behaves like the smooth min-entropy, and for large $\eps$ it approximates the smooth max-entropy.

\subsection{Basic properties}

\paragraph{Trivial bounds.}
For any state $\rho_{AB}$, the three entropy measures are lower-bounded  by 
$- \min \set{\log |A|, \log |B|}$, and  upper-bounded by $ \log |A|$.

\paragraph{Examples.}
For $\eps \to 0$, all three smooth entropies  are $0$ if $\rho_A$ is pure, $\log_2 \abs{A}$ if $\rho_{AB}  = \frac{1}{|A|} \id_A \otimes \rho_B$, and $-\log_2 \abs{A} $ if $\rho_{AB}$ is maximally entangled.

\paragraph{Pure bipartite states.}
The non-conditional versions of our entropies only depend on the spectrum of the reduced state, so, if $\rho_{AB}$ is pure, we have $\Hh^\eps (A) = \Hh^\eps(B)$, $\hmin^\eps (A) = \hmin^\eps(B)$ and $\hmax^\eps (A) = \hmax^\eps(B)$ (by Schmidt decomposition).

\paragraph{Conditioning on classical information.}
\cite[Prop.~4.6]{Tomamichel2012} 
For quantum-classical states of the form $\rho_{ABC} = \sum_k p_k \, \tau^k_{AB} \otimes \pure{k}_C$,  the conditional min- and max-entropies have the form 
\begin{align}
 \hmin(A|BC)_\rho &= - \log \left( \sum_k p_k 2^{- \hmin(A|B)_{\tau^k} } \right) ,  \label{eq:cqhmin}\\
 \hmax(A|BC)_\rho &= \log \left( \sum_k p_k 2^{ \hmax(A|B)_{\tau^k} } \right). \label{eq:cqhmax}\\
\end{align}

\paragraph{Product states.}
The conditional entropy equals the non-conditional entropy for product states, 
\begin{align}
\Hh^\eps (A|B)_{\rho_A \otimes \rho_B} = \Hh^\eps (A)_{\rho_A} .
\label{eq:products}
\end{align}
Equation~\ref{eq:products} also applies to the smooth min- and max-entropies.

\paragraph{Data-processing inequality.}
The entropy of $A$ conditioned on $B$ cannot decrease if information is  locally processed at $B$. Formally, 
\begin{align}
 \Hh^\eps(A |B)_{\rho} 
 \leq
 \Hh^\eps  (A| B')_{ [\I \otimes \E ] (\rho) },
 \label{eq:DPI}
\end{align}
where $[\I \otimes \E ] (\rho) $ is the state obtained from $\rho_{AB}$ after applying a trace-preserving completely positive map $\E$ on system $B$. 
Smooth entropies are invariant under local unitaries $U_A \otimes U_B$. This relation also holds for the smooth min- and max-entropies.

\subsection{Chain rules}

The hypothesis-testing entropy satisfies a chain rule.
\begin{lemma}[Cor.~1 from \cite{Dupuis2012}]\label{thm:chain_hh}
Let $\rho_{ABC}\in \End{A\otimes B\otimes C}$ be an arbitrary normalized state, and $\epsilon,\epsilon'>0$. Then,
\begin{align*}
 \Hh^{\epsilon+\sqrt{8\epsilon'}}(AB\vert C)_\rho\geq \Hh^\epsilon(A\vert BC)_\rho+H^{\epsilon'}(B\vert C)_\rho-\log\frac{\epsilon+\sqrt{8\epsilon'}}{\epsilon}.
\end{align*}

\end{lemma}

Smooth entropies satisfy several chain rules, for different combinations of min- and max-entropies~\cite{Vitanov2012}. 
Here we present those needed for our proofs.

\begin{lemma}[Lemma~A.7 from~\cite{Dupuis2010a}]\label{thm:chain_minmin}
 Let $\eps >0$ and $\eps', \eps''\geq 0$. Then
\begin{align*}
 \hmin^{\eps'}(A|BC)_\rho \leq \hmin^{\eps + 2 \eps ' + \eps ''} (AB|C)_\rho  -  \hmin^{\eps''}(B|C)_\rho + \log \frac1{1 - \sqrt{1- \eps^2} } .
\end{align*}
\end{lemma}

\begin{lemma}[Dual of Thm. 15 from~\cite{Vitanov2012}]  \label{thm:chain_maxmin}
Let $\eps >0$ and $\eps', \eps''\geq 0$. Then
\begin{align*}
 \hmax^{2\eps +  \eps ' + 2 \eps ''} (A|BC)_\rho  \leq     \hmax^{\eps'}(AB|C)_\rho -    \hmin^{\eps''}(B|C)_\rho +3 \log \frac1{1 - \sqrt{1- \eps^2} }.
\end{align*}
\end{lemma}

\begin{lemma}[Thm.~14 from~\cite{Vitanov2012}]  \label{thm:chain_minmax}
Let $\eps >0$ and $\eps', \eps''\geq 0$. Then
\begin{align*}
 \hmin^{\eps + \eps ' + \eps ''} (A|BC)_\rho \geq \hmin^{\eps'}(AB|C)_\rho - \hmax^{\eps''}(B|C)_\rho -  2 \log \frac1{1 - \sqrt{1- \eps^2} }.
\end{align*}
\end{lemma}

\begin{lemma}[Dual of Lemma~\ref{thm:chain_minmin}]  \label{thm:chain_maxmax}
Let $\eps >0$ and $\eps', \eps''\geq 0$. Then
\begin{align*}
 \hmax^{\eps''}(A|BC)_\rho \geq  \hmax^{\eps + 2 \eps ' + \eps ''} (AB|C)_\rho - \hmax^{\eps'}(B|C)_\rho - \log \frac1{1 - \sqrt{1- \eps^2} }.
\end{align*}
\end{lemma}

\subsection{Relations between the different smooth entropies}

\subsubsection{Duality between smooth min- and max-entropies}
For any tripartite pure state $\rho_{ABC}$, we have \cite{Konig2009,Tomamichel2010} 
 \begin{align}
  H_{\min}^\eps(A|C)_\rho = - H_{\max}^\eps(A|B)_{\rho} \ .
	\label{eq:duality}
\end{align}

\subsubsection{$\Hh^\eps$ interpolates between smooth min- and max-entropies}

\paragraph{ $H^\eps$ and $\hmin^{\eps'}$.}
For small $\eps$,  $H^\eps$ behaves approximately like the smooth min-entropy.
 \begin{align}
 \Hh^{\eps^2/2} (A| B)_\rho  
 &\leq \hmin^\eps   (A|B)_\rho 
 \leq  \Hh^{11\sqrt{\eps}} (A |B)_\rho
    + \frac52  \log \left(\frac3{\eps}  \right)
    + \log \left( \frac2{1-\eps}\right).
 \label{eq:relation_hhmin}  
\end{align}
The lower bound comes from \cite[Prop.~4.1]{Dupuis2012}. The upper bound is proved in Lemma~\ref{lemma:min_and_Hh}.

\paragraph{ $H^\eps$ and $\hmax^{\eps'}$.}
\cite[Prop.~8]{Dupuis2012} For large $\eps$, the  the hypothesis-testing entropy behaves approximately like max-entropy,
\begin{align}
 \hmax(A| B)_{\rho}+\log\frac{1}{\epsilon^2}\geq \Hh^{1-\epsilon}(A| B)_\rho.
\end{align}
There is also a known bound for the non-conditional smooth max-entropy, 
\begin{align}
  \Hh^{1-\epsilon}(A)_\rho \geq \hmax^{\sqrt{2\epsilon}}(A)_\rho + \log\frac{1}{(1-\epsilon)}.
 \label{eq:relation_hhmax}
\end{align}

\subsubsection{Smooth entropies and von Neumann entropy}
For a bipartite quantum state $\rho_{AB}$, the von Neumann entropy of $A$ conditioned on $B$ is defined as 
$H(A | B)_{\rho} = H(AB)_\rho - H(B)_\rho$, 
where $H(X)_\sigma = -\tr(\sigma_X \log_2 \sigma_X) $ is the usual (non-conditional) von Neumann entropy of $\sigma_X$.  
The conditional von Neumann entropy is always bounded by the smooth min- and max-entropies in the limit of small $\eps$ \cite{Tomamichel2009}, 
\begin{align} 
  \lim_{\eps \to 0} H_{\min}^\eps(A| B)_{\rho} 
	\leq H(A| B)_{\rho} 
	\leq \lim_{\eps \to 0} H_{\max}^\eps(A| B )_{\rho} 
	\label{eq:vNbound}
\end{align}
In particular, if the smooth min- and max-entropies coincide, they are automatically equal to the von Neumann entropy.

\paragraph*{Asymptotic equipartition property.}
Smooth entropy measures converge to the von Neumann entropy in the limit of many identical and independently distributed systems, when the global state has the form $\rho_{A^{\otimes n} B^{\otimes n}} = {\sigma_{AB}}^{\otimes n}$ \cite{Dupuis2012,Tomamichel2009}. Formally, for any $0 < \eps  <1$,
\begin{align} 
	  \lim_{n \to \infty} 
		\frac{1}{n} \Hh^{\eps}(A^{\otimes n} | B^{\otimes n} )_{\sigma^{\otimes n}}
	=\lim_{n \to \infty} 
		\frac{1}{n} \hmax^{\eps}(A^{\otimes n} | B^{\otimes n} )_{\sigma^{\otimes n}} 
	=  \lim_{n \to \infty} 
		\frac{1}{n} \hmin^{\eps}(A^{\otimes n} | B^{\otimes n} )_{\sigma^{\otimes n}}
	= H(A | B )_{\sigma} \ .
	\label{eq:AEP}
\end{align}

\newpage

\section{Decoupling theorems}
\label{appendix:decoupling}

Decoupling theorems~\cite{Dupuis2010, Dupuis2010a, Szehr2011} capture the idea that, given two quantum systems $A$ and $R$ not perfectly correlated, most (random) subsystems of $A$ up to a certain size are decoupled from $R$. The maximal size of decoupled subsystems depends on correlations between $A$ and $R$, as measured by conditional entropies. This result has powerful applications in quantum cryptography, error correction and thermodynamics \cite{Dupuis2010, DelRio2011, Szehr2011}.

\begin{thm}[{Decoupling [adapted from Thm. 3.1 of \cite{Dupuis2010a}]}]\label{thm:decoupling}
Let $\rho_{AR} \in \Set{A \otimes R }$. Let $\cT_{A \to B}$ be a trace non-increasing, completely positive map from $\End{A}$ to $\End{B}$.
Let $\tau$ be the  \cj representation  of $\cT$,
$$\tau_{A'B}= [\I_{A'} \otimes \cT_{A \to B} ] \ \left( \pure{\Psi}_{A'A}\right) ,$$
where $ \ket{\Psi}_{A'A} = \abs A^{- \frac12} \sum_i^{\abs A} \ket i_A \ket i_{A'}$ is maximally entangled  between $A'$ and a virtual system $A$.
Finally, let $\eps, \Delta, \delta >0$.

If the entropic relation
\begin{align*}
   \hmin^\eps (A|R)_\rho + \hmin^\eps (A'|B)_\tau \geq  2 \log \ \frac1{ \Delta - 12 \eps} ,
\end{align*}
holds,
then the fraction (over the set of all unitaries $\set{U_A}$ acting on $A$, according to the Haar measure) of unitaries such that
\begin{align*} 
  \|  [\cT \otimes \I_R]   (  U_A \cdot \rho_{A R}  )- \tau_B \otimes \rho_R  \|_1
 \geq  
    \Delta
    + \delta
\end{align*}
is at most $  2\ e^{- \frac{ \abs{A}}{ 16}  \delta^2}$.
\end{thm}

Note that $\delta, \Delta$ and $\eps$ do not scale with the size of the systems involved, whereas the entropies do. 

The converse theorem gives us tightness of the bound above for trace-preserving maps.

\begin{thm}[Converse] \label{thm:converse}
Let  $\rho_{AR} \in \Set{A \otimes R }$. 
Let $\cT_{A \to B}$ be a trace-preserving completely positive map from $\End{A}$ to $\End{B}$.
Let
$$\tilde\rho_{BA'} := [ \cT_{A \to B} \otimes   \I_{A'} ] (\rho_{AA'}), $$
where $\rho_{AA'}$ is a purification of $\rho_A = \tr_R ( \rho_{AR})$ on a virtual system $A'$. 
For any $\eps' > 0$ and any $\eps, \eps'' \geq 0$, if
\begin{align}
 \hmin^{2 \sqrt{ 2 \eps+6\eps''} + 2 \sqrt{\eps'} +\eps''}(A|R)_\rho + \hmax^{\eps''} (A'|B)_{\tilde\rho} 
 < - \log  \frac{1}{\eps'},
 \label{eq:converse_condition}
\end{align}
then
\begin{align*}
 \| [ \cT \otimes \I_R]  (\rho_{AR}) - \cT (\rho_A) \otimes \rho_R \| > \eps .
\end{align*}
\end{thm}

The following corollary is useful to compare the final state with the canonical state.

\begin{corollary} \label{cor:converse}
 In the setting of Thm.~\ref{thm:converse}, if condition (\ref{eq:converse_condition}) holds, then
 \begin{align*}
    \| [ \cT \otimes \I_R]  (\rho_{AR}) - \cT (\sigma_A) \otimes \rho_R \| > \frac\eps2 ,
 \end{align*}
 for any normalized density operator $\sigma_A$ on $A$.
\end{corollary}

\begin{proof}
First we use the fact that the trace distance cannot decrease under trace-preserving completely positive maps, like the partial trace, to show
\begin{align*}
 \| [ \cT \otimes \I_R]  (\rho_{AR}) - \cT (\sigma_A)  \otimes \rho_R \| 
% &\geq \| \tr_R \left(  [ \cT \otimes \I_R]  (\rho_{AR}) \right) - \tr_R \left( \cT (\sigma_A)  \otimes \rho_R \right)\| \\
 &\geq \|  \cT  (\rho_{A}) -  \cT (\sigma_A)  \| \\
 &= \|  \cT (\rho_A) \otimes \rho_R - \cT (\sigma_A)   \otimes \rho_R \| .
\end{align*}
Using the triangle inequality for the trace distance, we obtain
\begin{align*}
 \eps 
  &<    \| [ \cT \otimes \I_R]  (\rho_{AR}) - \cT (\rho_A) \otimes \rho_R \|  \\
  &< \| [ \cT \otimes \I_R]  (\rho_{AR}) - \cT (\sigma_A)  \otimes \rho_R \|  
   + \|  \cT (\sigma_A)   \otimes \rho_R - \cT (\rho_A) \otimes \rho_R \| \\
  &< 2\ \| [ \cT \otimes \I_R]  (\rho_{AR}) - \cT (\sigma_A)  \otimes \rho_R \|  .
\end{align*}

\end{proof}

\newpage

\section{Detailed results and proofs}
\label{appendix:proof}

\emph{Lasciate ogni speranza, voi ch'entrate.}

\subsection{Thermalization of typical subsystems}
\label{appendix:proof_direct}

In this section we prove our main result on thermalization after a random evolution (or thermalization of random subsystems), Thm.~\ref{thm:typical_main}.
The first step is to apply  the  decoupling theorem (Thm.~\ref{thm:decoupling}), setting $A = \Omega$, $B= S$, and $\cT_{\Omega \to S} = \tr_E$.
\begin{lemma} \label{lemma:decoupling_trace}

Let $\rho_{SER} \in \Set{ \Omega\otimes R}$, with $\Omega \subseteq S\otimes E $.    
For any $\tilde\eps \geq0$, and any $\Delta >0$, if
\begin{align}
   \hmin^{\tilde\eps} (\Omega|R)_\rho + \hmin^{\tilde\eps} (\Omega'|S)_\tau \geq - 2 \log( \Delta - 12 \tilde\eps)
   \label{eq:decoupling_condition_trace}
\end{align}
holds, then, for any $\delta>0$, the fraction of unitaries $\set{U_\Omega}$ acting on $\Omega$  such that
\begin{align*} 
  \| \tr_E   (  U_\Omega \cdot \rho_{\Omega R}  )- \pi_S \otimes \rho_R  \|_1
 \geq  
    \Delta
    + \delta
\end{align*}
is at most $  2\ e^{- \frac{ \abs{\Omega}}{ 16}  \delta^2}$, according to the Haar measure.

In the above,
$ \tau_{\Omega' S} = \tr_E (\pure{\Psi}_{\Omega' \Omega} )$, for the maximally entangled state $\ket\Psi_{\Omega' \Omega}$. 
Note that the reduced state in $S$ is the canonical state, $\tau_S =\tr_{\Omega'} \tr_E (\pure{\Psi}_{\Omega' \Omega}) = \pi_S$.
\end{lemma}

Now we are ready to state our main theorem in terms of the smooth min- and max-entropies. 
A final reformulation in terms of $\Hh^\eps$ follows (Corollary~\ref{cor:typical}). 

\begin{thm}\label{thm:main}
Let $\rho_{SER} \in \Set{ \Omega\otimes R}$, with $\Omega \subseteq S\otimes E $.   
For any $ \eps_2, \eps_3 \geq 0$, any $\eps_1 > \eps_2 + \eps_3 $, and any  $\Delta > 0$, if the entropic relation
\begin{align*}
   \hmin^{\eps_1} (SE|R)_\rho +     \hmin^{\eps_2}(E)_\pi   - \hmax^{\eps_3 }(S)_\pi 
   \geq  2 \log \frac1{(1 - \sqrt{1- (\eps_1 - \eps_2 - \eps_3)^2 } ) ( \Delta - 12 \eps_1 ) }
\end{align*}
 holds, then, for any $\delta>0$, the fraction  of unitaries  $\set{U_\Omega}$ acting on $\Omega$  such that
\begin{align*} 
  \| \tr_E   (  U_\Omega \cdot \rho_{\Omega R}  )- \pi_S \otimes \rho_R  \|_1
 \geq  
    \Delta
    + \delta
\end{align*}
is at most $  2\ e^{- \frac{ \abs{\Omega}}{ 16}  \delta^2}$, according to the Haar measure.
\end{thm}

\begin{proof}

We start from Lemma~\ref{lemma:decoupling_trace}, and break down the left-hand side of condition (\ref{eq:decoupling_condition_trace}).  First off, we observe that $\hmin^{\tilde \eps} (\Omega|R)_\rho = \hmin^{\tilde \eps} (SE|R)_\rho$.
We use the chain rule from Lemma~\ref{thm:chain_minmax}  to bound the other entropy. Setting $\tilde{\eps} = \eps_1 + \eps_2 + \eps_3$, we have
\begin{align*}
 \hmin^{\eps_1 + \eps_2 +\eps_3} (\Omega'|S)_\tau 
   \geq 
   \hmin^{\eps_2 } (\Omega ' S)_\tau 
   - \hmax^{\eps_3 }(S)_\tau 
   + 2 \log  \left(1 - \sqrt{1- \eps_1^2} \right)  .
\end{align*}
Since $\ket{\Psi}_{\Omega' S E}$ is a pure state, we have that $\hmin^{\eps_2}(\Omega' S)_\tau  = \hmin^{\eps_2}(E)_\pi$.
Condition (\ref{eq:decoupling_condition_trace}) becomes
\begin{align*}
   \hmin^{\eps_1 + \eps_2 + \eps_3} (SE|R)_\rho +     \hmin^{\eps_2}(E)_\pi   - \hmax^{\eps_3 }(S)_\pi 
   + 2 \log  \left(1 - \sqrt{1- \eps_1^2} \right)     \geq - 2 \log( \Delta - \eps_1 - \eps_2 -\eps_3 ).
\end{align*}
To clean up, we take $\eps_1 + \eps_2 +\eps_3 \to \eps_1$.

\end{proof}

We may now write this result in terms of the hypothesis-testing entropy, and simplify the $\eps$ terms at the cost of little generality. 

\begin{corollary} \label{cor:typical}
 Let $\rho_{SER} \in \Set{ \Omega\otimes R}$, with $\Omega \subseteq S\otimes E $.   
Let  $\eps, \Delta > 0$.

If the entropic relation
\begin{align*}
 \Hh^{ 9   \eps } (SE|R)_\rho 
     + \Hh^{\eps}(E)_\pi   
     -  \Hh^{1- \eps}(S)_\pi 
 \geq  2 \log \frac1{   (1 - \sqrt{1- 2 \ \eps }) ( \Delta - 36 \ \sqrt{2 \eps}  ) }
    - \log\frac1{1-\eps}
\end{align*}
 holds, then, for any $\delta>0$, the fraction  of unitaries  $\set{U_\Omega}$ acting on $\Omega$  such that
\begin{align*} 
  \| \tr_E   (  U_\Omega \cdot \rho_{\Omega R}  )- \pi_S \otimes \rho_R  \|_1
 \geq  
    \Delta
    + \delta
\end{align*}
is at most $  2\ e^{- \frac{ \abs{\Omega}}{ 16}  \delta^2}$, according to the Haar measure.
\end{corollary}

\begin{proof}
Starting from 
\begin{align}
   \hmin^{\eps_1} (SE|R)_\rho +     \hmin^{\eps_2}(E)_\pi   - \hmax^{\eps_3 }(S)_\pi 
   \geq  2 \log \frac1{(1 - \sqrt{1- (\eps_1 - \eps_2 - \eps_3)^2 } ) ( \Delta - 12 \eps_1 ) },
   \label{eq:whatever}
\end{align}
we use relations (\ref{eq:relation_hhmin}) and (\ref{eq:relation_hhmax}) to obtain
\begin{align*}
   & \hmin^{\eps_1} (SE|R)_\rho \geq \Hh^{\frac{{\eps_1}^2}2} (SE|R)_\rho, \\
   & \hmin^{\eps_2}(E)_\pi  \geq \Hh^{\frac{{\eps_2}^2}2}(E)_\pi, \\
   & -\hmax^{\eps_3 }(S)_\pi \geq -  \Hh^{1- \frac{{\eps_3}^2}2}(S)_\pi + \log\frac{1}{1-{\eps_3}^2 / 2}.
\end{align*}
Applying these bounds to (\ref{eq:whatever}), we get 
\begin{align*}
   \Hh^{\frac{{\eps_1}^2}2} (SE|R)_\rho 
    + \Hh^{\frac{{\eps_2}^2}2}(E)_\pi   
    -  \Hh^{1- \frac{{\eps_3}^2}2}(S)_\pi 
   \geq  2 \log \frac1{(1 - \sqrt{1- (\eps_1 - \eps_2 - \eps_3)^2 } ) ( \Delta - 12 \eps_1 ) } - \log\frac{1}{1-{\eps_3}^2 / 2}.
\end{align*}
To simplify, we consider the special case $\tilde\eps = \frac{\eps_1}3 = \eps_2 = \eps_3$. This gives us
\begin{align*}
   \Hh^{\frac{9  \tilde\eps^2}2} (SE|R)_\rho 
    + \Hh^{\frac{\tilde\eps^2}2}(E)_\pi   
    -  \Hh^{1- \frac{\tilde\eps^2}2}(S)_\pi 
   \geq  2 \log \frac1{(1 - \sqrt{1- \tilde\eps^2 } ) ( \Delta - 36 \ \tilde\eps ) } - \log\frac{1}{1-{\tilde\eps}^2 / 2}.
\end{align*}
Finally, we  take $\eps = \frac{\tilde\eps^2}2$ to obtain the statement of the corollary. 
\end{proof}

The result presented in the main part of this work is obtained by taking $\Delta = \delta$.

% xxxxxxxxxxxxxxxxxxxxxxxxxxxxxxxxxxxxxxxxxxxxxxxxxxxxxxxxxxxxxxxxxxxxxxxxxxxxxxxxxxxxxxxxxxxxxxxxxxxxxx
\subsection{Converse}
\label{appendix:converse}

The converse bound follows. A friendlier, if weaker, bound can be found in Corollary \ref{cor:typConverse}. 

\begin{thm}[Tightness]\label{thm:typConverse}

  Let $\rho_{SER} \in \Set{ \Omega\otimes R}$, with $\Omega \subseteq S\otimes E $.   
  Let $\delta, \eps_1, \eps_2 > 0$ and $ \eps_3, \eps_4\geq 0$.
  
  For readability, we set $\tilde\eps = 2 \sqrt{ \delta+3(\eps_2 +  \eps_3 + \eps_4)} + \eps_1 + \eps_2 +  \eps_3 +  \eps_4$.
  
  If
  \begin{align}
   \hmin^{2 \tilde\eps  }(\Omega|R)_\rho + \max_{\sigma \in \Set\Omega}  \ [ \hmax^{2 \eps_3 } (E)_\sigma  -  \hmin^{\eps_4}(S)_\sigma ] 
  < - \log  \frac{1}{{\eps_1}^2} - 3 \log \frac1{1 - \sqrt{1- {\eps_2}^2} },
  \label{eq:converse_unitary_condition}
 \end{align}
  then
  \begin{align*}
  \|  \tr_E ( U_\Omega \cdot \rho_{AR}) - \pi_S \otimes \rho_R \| >\delta,
  \end{align*} 
  for any unitary $U_\Omega$ acting on $\Omega$.
\end{thm}

\begin{proof}
 We start from Cor.~\ref{cor:converse}, setting $A=\Omega$, $B=S$, $\cT (\cdot) = \tr_E(U_\Omega \cdot ) $, and $\sigma_A = \pi_\Omega$. This gives us the condition 
   \begin{align} \label{eq:first_condition_converse_subsystems}
  \hmin^{2 \sqrt{ 2 \eps_1+6\eps_2} + 2 \sqrt{\eps_3} +\eps_2}(\Omega|R)_\rho + \hmax^{\eps_2} (\Omega'|S)_{\tilde\rho} 
  < - \log  \frac{1}{\eps_3},
  \end{align}
which implies
  \begin{align*}
  \|  \tr_E (U_\Omega \cdot \rho_{AR}) - \tr_E (\pi_\Omega) \otimes \rho_R \| > \frac{\eps_1}2 .
  \end{align*}
Here, $\tilde\rho = U_\Omega \cdot  \rho_{\Omega \Omega'}$, where $\rho_{\Omega \Omega'}$ is a purification of $\rho_\Omega$.

We will look for an upper bound for $\hmax^{\eps_2} (\Omega'|S)_{\tilde\rho} $, as we might not know which unitary $U_\Omega$ was applied. We will use a little of brute force, maximizing the conditional entropy over all states $\sigma_\Omega$ in $\Set{\Omega}$, with purification $\sigma_{\Omega\Omega'}$ (this is stronger than maximizing over all unitaries $U_\Omega$). 
Also, in order to use a chain rule, let us set $\eps_2 = 2\eps_4 +  \eps_5 + 2 \eps_6$. We have
\begin{align*}
  \hmax^{2\eps_4 +  \eps_5 + 2 \eps_6} (\Omega'|S)_{\tilde\rho} 
    &\leq \max_{\sigma \in \Set\Omega}  \hmax^{2\eps_4 +  \eps_5 + 2 \eps_6} (\Omega'|S)_{\sigma_{\Omega\Omega'}} \\
    &\leq \max_{\sigma \in \Set\Omega}  \ [ \hmax^{\eps_5} (S\Omega')_\sigma  -  \hmin^{\eps_6}(S)_\sigma ] +3 \log \frac1{1 - \sqrt{1- \eps_4^2} } \qquad \text{\flag{Lemma~\ref{thm:chain_maxmin}}} \\
    &= \max_{\sigma \in \Set\Omega}  \ [ \hmax^{\eps_5} (E)_\sigma  -  \hmin^{\eps_6}(S)_\sigma ] +3 \log \frac1{1 - \sqrt{1- \eps_4^2} } \qquad  \flag{ \sigma_{SE\Omega'}\text{ pure}} .
\end{align*}    
Condition (\ref{eq:first_condition_converse_subsystems}) becomes
\begin{align*}
   \hmin^{2 \sqrt{ 2 \eps_1+6(2\eps_4 +  \eps_5 + 2 \eps_6)} + 2 \sqrt{\eps_3} + (2\eps_4 +  \eps_5 + 2 \eps_6)}(\Omega|R)_\rho + \max_{\sigma \in \Set\Omega}  \ [ \hmax^{\eps_5} (E)_\sigma  -  \hmin^{\eps_6}(S)_\sigma ] \\
  < - \log  \frac{1}{\eps_3} - 3 \log \frac1{1 - \sqrt{1- \eps_4^2} },
\end{align*}
which we cannot hope to make much more readable without losing generality (we do simplify it in the corollary ahead). For now, let us just relabel 
$$\eps_1 \to 2 \delta, \quad  \eps_3 \to \eps_1^2, \quad \eps_4 \to \eps_2, \quad  \eps_5 \to 2 \eps_3, \quad \eps_6 \to \eps_4 ,$$ 
to obtain the beauty
\begin{align*}
   \hmin^{2 \left( 2 \sqrt{ \delta+3(\eps_2 +  \eps_3 + \eps_4)} + \eps_1 + \eps_2 +  \eps_3 +  \eps_4  \right)   }(\Omega|R)_\rho + \max_{\sigma \in \Set\Omega}  \ [ \hmax^{2 \eps_3 } (E)_\sigma  -  \hmin^{\eps_4}(S)_\sigma ] \\
  < - \log  \frac{1}{{\eps_1}^2} - 3 \log \frac1{1 - \sqrt{1- {\eps_2}^2} }.
\end{align*}

\end{proof}

In the following corollary we simplify some of the terms. In particular, we neglect a term with the smooth min-entropy of $S$, for an optimal state. In the typical case where $S$ is much smaller than $E$, this is only a small loss. 

\begin{corollary} \label{cor:typConverse}
  Let $\rho_{SER} \in \Set{ \Omega\otimes R}$, with $\Omega \subseteq S\otimes E $.  
  Let $\delta> 0$ and let $\eps> 4 \sqrt{\delta}$. 
  For simplicity,\footnote{That was a joke. You are smiling too.} we define $f(\eps, \delta) := \frac1{16}(6+\eps -2 \sqrt{9+3\, \eps + 4 \delta})^2$. 
  
  If
  \begin{align}
  \Hh^{11\sqrt{ \eps}} (\Omega |R)_\rho + \Hh^1(E)_\pi 
  < - \log  \frac{1}{  f(\eps,\delta) } - 3 \log \frac1{1 - \sqrt{1- f(\eps, \delta) }}
    - \frac52  \log \left(\frac3 \eps  \right)
    - \log \left( \frac2{1- \eps}\right),
  \label{eq:converse_unitary_condition_cor}
 \end{align}
 then 
 \begin{align*}
  \|  \tr_E ( U_\Omega \cdot \rho_{AR}) - \pi_S \otimes \rho_R \| > \delta ,
  \end{align*} 
  for any unitary $U_\Omega$ acting on $\Omega$.
\end{corollary}

\begin{proof}

We start from the condition of Thm.~\ref{thm:typConverse},
\begin{align*}
   \hmin^{2 \left( 2 \sqrt{ \delta+3(\eps_2 +  \eps_3 + \eps_4)} + \eps_1 + \eps_2 +  \eps_3 +  \eps_4  \right)   }(\Omega|R)_\rho + \max_{\sigma \in \Set\Omega}  \ [ \hmax^{2 \eps_3 } (E)_\sigma  -  \hmin^{\eps_4}(S)_\sigma ] \\
  < - \log  \frac{1}{{\eps_1}^2} - 3 \log \frac1{1 - \sqrt{1- {\eps_2}^2} }.
\end{align*}
We are looking for a simpler, tighter condition, \ie an upper bound to the left-hand side of the inequality and a lower bound to the right-hand side.\footnote{In other words, we start from an inequality like $A < B$, and search for good-looking $\bar A$ and $\bar B$ such that $A \leq \bar A$ and $\bar B \leq B$. Therefore, $\bar A < \bar B$ implies the original condition $A < B$. }
First we neglect the term with the non-conditional entropy of $S$, as 
\begin{align*}
 \max_{\sigma \in \Set\Omega}  \ [ \hmax^{\eps_3} (E)_\sigma  -  \hmin^{\eps_4}(S)_\sigma ] 
 &\leq \max_{\sigma \in \Set\Omega}  \  \hmax^{\eps_3} (E)_\sigma .
\end{align*}
Now we apply the upper bound for the max-entropy given by Lemma~\ref{lemma:upperbound_hmax},
\begin{align}
 \max_{\sigma \in \Set\Omega}  \  \hmax^{\eps_3} (E)_\sigma  
 \leq \max_{\sigma \in \Set\Omega}  \ \Hh^1(E)_\sigma
 =\max_{\sigma \in \Set\Omega} \  \log |\text{supp } \sigma_E |.
 \label{eq:boundSupport}
\end{align}
Finally, we show that for all states  $\sigma \in \Set\Omega$, it stands that
$\text{supp } \sigma_E \subseteq \text{supp } \pi_E$, and therefore (\ref{eq:boundSupport}) is upper-bounded by 
$\Hh^1(E)_\pi$.
For every $\sigma \in \Set{\Omega}$, there exists a basis $\set{\ket k}_k$ of $\Omega$ that diagonalizes it,
\begin{align*}
 \sigma = \sum_k^{|\Omega|} p_k \pure k_\Omega.
\end{align*}
Since $\Omega$ is a subspace of $S\otimes E$, we can expand each  element $\ket k_\Omega$ in any basis of $S\otimes E$; in particular, we can choose a product basis $\set{\ket i_S \otimes \ket j_E}_{i,j}$, such that
\begin{align*}
 \ket k_\Omega &= \sum_i^{|S|} \sum_j^{|E|} c_{ij}^k \ \ket i_S \otimes \ket j_E, \qquad
   \sum_{i,j} |c_{ij}^k|^2 = 1, \ \forall\ k.
\end{align*}
We may now expand $\sigma$ in this basis,
\begin{align*}
  \sigma_\Omega &= \sum_k^{|\Omega|} p_k \sum_{i, i'}^{|S|} \sum_{j, j'}^{|E|}  c_{ij}^k \ {(c_{i'j'}^k)}^* \  \proj{i}{i'}_S \otimes \proj{j}{j'}_E, \qquad
  \sigma_E = \tr_S  \ \bar\sigma_\Omega = \sum_k^{|\Omega|} p_k \sum_{i}^{|S|} \sum_{j, j'}^{|E|}  c_{ij}^k \ {(c_{ij'}^k)}^* \  \proj{j}{j'}_E.
\end{align*}
Note that the canonical state is given by 
\begin{align*}
  \pi_\Omega &= \sum_k^{|\Omega|} \frac1{|\Omega|} \pure k \\
  &= \sum_k^{|\Omega|} \frac1{|\Omega|}  \sum_{i, i'}^{|S|} \sum_{j, j'}^{|E|}  c_{ij}^k \ {(c_{i'j'}^k)}^* \  \proj{i}{i'}_S \otimes \proj{j}{j'}_E, \qquad
  \pi_E = \sum_k^{|\Omega|} \frac1{|\Omega|}  \sum_{i}^{|S|} \sum_{j, j'}^{|E|}  c_{ij}^k \ {(c_{ij'}^k)}^* \  \proj{j}{j'}_E,
\end{align*}
so clearly $\text{supp } \sigma_E \subseteq \text{supp } \pi_E$. 

Let us see where we stand. We may set $\eps_3= \eps_4 = 0$, and $\eps_1=\eps_2=:\tilde \eps$. Our condition becomes
\begin{align*}
  \hmin^{4 \left(\sqrt{ \delta+3\tilde\eps }\  + \tilde \eps  \right)   }(\Omega|R)_\rho +  \Hh^1(E)_\pi 
  < - \log  \frac{1}{{\tilde\eps}^2} - 3 \log \frac1{1 - \sqrt{1- {\tilde \eps}^2} }
\end{align*}
We may also bound the term with the smooth min-entropy using  \eref{eq:relation_hhmin}. We set $\eps := 4 (\sqrt{ \delta+3\tilde \eps }  + \tilde \eps)$, and have
\begin{align*}
 \hmin^{ \eps  }(\Omega|R)_\rho
 &\leq \Hh^{11\sqrt{ \eps}} (\Omega |R)_\rho
    - \frac52  \log \left(\frac{ \eps}3   \right)
    + \log \left( \frac2{1- \eps}\right).
\end{align*}
This leaves us with the condition
 \begin{align*}
  \Hh^{11\sqrt{ \eps}} (\Omega |R)_\rho + \Hh^1(E)_\pi 
  < - \log  \frac{1}{\tilde\eps^2} - 3 \log \frac1{1 - \sqrt{1- {\tilde\eps}^2} }
    - \frac52  \log \left(\frac3{ \eps}  \right)
    - \log \left( \frac2{1- \eps}\right) .
 \end{align*}
Now we should make the dependence in $\delta$ a little more explicit. In order to keep the above expression only moderately foul, we bound the logarithmic terms on the right-hand side. 
We shall spare you the details (but if you insist, we used ${\tilde\eps}^2 = \frac1{16}(6+\eps -2 \sqrt{9+3\, \eps + 4 \delta})^2$, applied the expansion $1-\sqrt{1-x^2} \geq \frac{x^2}2$ twice, and sacrificed a black chicken).
% We thank the beta reader who reminded us of the twelve idols aligned in a circle around a fire at midnight under the first full moon of winter. If only we had tried that first.
The new bound is
 \begin{align*}
  \Hh^{11\sqrt{ \eps}} (\Omega |R)_\rho + \Hh^1(E)_\pi 
  < - 4 \log  \frac{6+\eps}{  4(\frac{\eps^2}{16} - \delta)^2 } 
    - \frac52  \log \left(\frac3 \eps  \right)
    - \log \left( \frac{16}{1- \eps}\right) .
 \end{align*}
 
\end{proof}

\subsection{Dimension bounds}
\label{appendix:dimensions}

To give an intuitive idea of the magnitude of the entropic terms in our results, we present a coarser version of our direct bounds.

\begin{corollary}
Let $\rho_{SER} \in \End{ \Omega\otimes R}$ be a normalized density operator, with $\Omega \subseteq S\otimes E $.    
For any $\eps \geq0$, and any $\Delta >0$, if
\begin{align}
   \Hh^{\eps} (\Omega|R)_\rho +  \log |\Omega| - 2 \log |S| \geq  2 \log\left( \frac1{\delta -  \sqrt{2\,\eps}} \right)
\end{align}
holds, then, for any $\delta>0$, the fraction of unitaries $\set{U_\Omega}$ acting on $\Omega$  such that
\begin{align*} 
  \frac12 \ \| \tr_E   (  U_\Omega \cdot \rho_{\Omega R}  )- \pi_S \otimes \rho_R  \|_1
 \geq  
     \delta
\end{align*}
is at most $  2\ e^{- \frac{ \abs{\Omega}}{ 16}  \delta^2}$, according to the Haar measure.
\end{corollary}

This corollary follows directly from Lemma~\ref{lemma:decoupling_trace}, combined with Lemma~\ref{thm:hmindimensions}, and the relation between the smooth-min entropy and the hypothesis-testing entropy. We also set $\Delta = \delta$.

\newpage

\section{A profusion of little lemmas for smooth entropies}
\label{appendix:lemmas}

In order to prove our physical results, we needed to show some properties of smooth entropies. 
This appendix is a collection of technical lemmas, mostly adaptations of similar results for other entropy measures. Please have no expectations of elegance or originality as you read through. 

The highlights of the appendix are Lemma~\ref{lemma:smoothness_Hh}, where we show that $\Hh^\eps$ is continuous on the quantum state (in a way that does not depend on the dimension of the quantum systems involved; in other words, it is ``smooth''), and Lemma~\ref{lemma:min_and_Hh}, where we give a bound for $\Hh^\eps$ in terms of the conditional smooth min-entropy.

\subsection{A few more definitions}

\subsubsection{Hypothesis-testing relative entropy  as a semi-definite program}

We can write the hypothesis-testing relative entropy as a semi-definite program (SDP) \cite{Boyd2004, Watrous2009, Dupuis2012}. The primal and dual SDPs for  $2^{-D^\eps (\rho||\sigma)}$ are

\begin{center}
\begin{tabular}{r  l}
\textsc{Primal}\\
 \quad \\
 minimize \quad
   & $\frac1\eps \tr (Q \,  \sigma)$ \\
 \quad \\
 subject to \quad
   & $\tr (Q \,  \rho) \geq \eps$,\\
   &$0 \leq Q \leq \id$ 
\end{tabular} 
\qquad \qquad
\begin{tabular}{r  l}
\textsc{Dual}\\
 \quad \\
 maximize \quad
   & $\mu - \frac{\tr X}{\eps}$\\
 \quad \\
 subject to \quad
   &$\mu\, \rho \leq \sigma + X$,\\
   &$X, \mu \geq 0$.
\end{tabular} 
\end{center}
In the above, it is required that $\rho$ and $\sigma$ be Hermitian operators. Remember that the generalized conditional smooth entropy is defined as $H^\eps(A|B)_\rho = - D^\eps(\rho_{AB}|| \id_A \otimes \rho_B)$.

\subsubsection{Alternative smooth min-entropy}

$\hat H_{\min}^\eps$ is an alternative entropy measure similar to the smooth min-entropy, except that we do not optimize over the choice of the marginal $\sigma_B$ \cite{Renner2005}. 
\begin{align*}
  \hat H_{\min}^\eps (A|B)_{\rho} := \max_{\tilde \rho_{AB} \in \mathcal B^\eps(\rho) }  \ 
  \sup_{\lambda \in \mathbb{R}} \set{ \lambda:  2^{- \lambda} \id_A \otimes \tilde \rho_B \geq \tilde \rho_{AB} }.
\end{align*} 
The optimization is made over the set of subnormalized states that are $\eps$-close to $\rho_{AB}$, according to the purified distance.

\subsection{A couple of trivial bounds for the smooth entropies}

\begin{lemma} \label{lemma:H1}
 Let $\rho \in \Set{A}$. Then we have
 $\Hh^1(A)_\rho = \log \abs {\text{supp } \rho } $
\end{lemma}

\begin{proof}
To show that $\Hh^1(A)_\rho \leq \log \abs {\text{supp } \rho } $, we look at the primal program for $\Hh^1(A)_\rho$,
\begin{align*}
 2^{\Hh^1(A)_\rho} &= \min \tr (Q_A \, \id_A), \\
\tr (Q_A \, \rho_A ) &\geq 1, \quad 0 \leq Q_A \leq \id_A. 
\end{align*}
We take as a candidate the projector onto the support of $\rho$, $Q= \Pi_\rho$. We have $\tr (\Pi_\rho \, \rho ) =1$, so $\Pi_\rho$ is a feasible candidate for the minimization. Therefore we have $2^{\Hh^1(A)_\rho} \leq \tr (\Pi_\rho \, \id_A) = \abs {\text{supp } \rho } $.

Now we show that $\Hh^1(A)_\rho \geq \log \abs {\text{supp } \rho } $. The dual program for the generalized smooth entropy $\Hh^1(A)_\rho$ is, in the non-conditional case, 
\begin{align*}
 2^{\Hh^1(A)_\rho} &= \max \mu - \tr X, \\
 \mu \rho &\leq \id + X, \quad \mu, X \geq 0. 
\end{align*}
Let us take the candidate $X = \mu \rho - \Pi_\rho$. We have 
\begin{align*}
  \id + X 
  = \id + \mu \rho - \Pi_\rho \geq \mu \rho,
\end{align*}
so $X$ is a feasible candidate for the dual SDP. This gives us
\begin{align*}
  2^{\Hh^1(A)_\rho} 
  &\geq \mu - \tr X 
  = \mu - \tr (\mu \rho - \Pi_\rho ) 
  = \mu - \mu \tr \rho + \tr (\Pi_\rho )
  = \abs {\text{supp } \rho }. 
\end{align*}
\end{proof}

\begin{lemma} \label{lemma:upperbound_hmax}
Let $\rho \in \Set{A}$. The non-conditional max-entropy is upper bounded as 
\begin{align*}
 \hmax^\eps(A)_\rho \leq \Hh^1(A)_\rho.
\end{align*}

\end{lemma}

\begin{proof}
Let $\rho_A = \sum_k p_k \ \pure{k}_A$, for some basis $\set{\ket k}_k$ of the support of $\rho$ in $A$. We note that $\hmax^\eps(A)_\rho  \leq \hmax^0 (A)_\rho = \log_2 F(\rho_{A}, \id_{A} )^2$, and
\begin{align*}
 F\left(\rho_A, \id_A \right)^2
 &= \tr {\left( \left|\sqrt{\rho_A} \ \sqrt{\id_A} \right|\right)}^2 \\
%  &= \tr \left| \sum_{k}^{\text{supp } \rho} \sqrt{p_k} \ \pure{k}_A \right|^2 \\
% &= \left(\sum_{k}^{|\text{supp } \rho|} \sqrt{ p_k} \right)^2\\
 &= \sum_{k,\ell}^{|\text{supp } \rho|}  \sqrt{ p_k}  \sqrt{ p_\ell} \\
 &\leq \sum_{k,\ell}^{|\text{supp } \rho|}   \frac{p_k + p_\ell}2  \qquad \flag{\text{inequality of arithmetic and geometric means}} \\
 &= |\text{supp } \rho_A| .
\end{align*}
Combining this with Lemma~\ref{lemma:H1}, we obtain $ \hmax^\eps(A)_\rho \leq \Hh^1(A)_\rho$.
\end{proof}

The following lemma is used to bound our condition for relative thermalization in terms of system dimensions (see Appendix~\ref{appendix:dimensions}).

\begin{lemma}\label{thm:hmindimensions}
Let $\rho_{AB} \in \Set{A \otimes B}$ be a quantum state with a fully mixed marginal in $A$,  $\rho_A = \frac{\id_A}{|A|}$.
Then, for any $\eps \geq 0$, 
\begin{align*}
 \hmin^\eps(A|B)_\rho \geq \log  |A|  - 2 \log |B| .
\end{align*}
\end{lemma}

\begin{proof}
We start by going to the non-smooth version of the min-entropy, 
\begin{align*}
 \forall \eps \geq 0, \qquad   \hmin^\eps(A|B)_\rho \geq  \hmin (A|B)_\rho. 
\end{align*}
It is convenient to formulate the min-entropy as an SDP. The primal SDP for $2^{- \hmin (A|B)_\rho}$ is

\begin{center}
\begin{tabular}{r  l}
 minimize \quad
   & $\gamma$ \\
 \quad \\
 subject to \quad
   & $\rho_{AB} \leq \gamma  \ \id_A \otimes \sigma_B$,\\
   & $ \sigma_B \in \Set{B}$,  \\
   &$\gamma \geq 0.$ 
\end{tabular} 
\end{center}

We want to show that $\gamma = \frac{|B|^2}{|A|}$ is a feasible candidate for the optimization problem, so that  $\hmin (A|B)_\rho \geq  \log \frac{|A|}{|B|^2}$. 
We apply  \cite[Lemma A.2]{Tomamichel2012}, which says that for positive operators $\rho \in \End{A\otimes B}$, it holds that $ \rho_{AB} \leq |B| \ \rho_A \otimes \id_B$. This gives us
\begin{align*}
 \rho_{AB} 
   &\leq |B| \ \rho_A \otimes \id_B  
   = |B| \  \frac{\id_A}{|A|}\otimes \id_B 
   = \frac{|B|^2}{|A|} \ \id_A \otimes \underbrace{\frac{\id_B}{|B|}}_{\green{=: \sigma_B}} .
\end{align*}

% <3 Brunch, Lea.

\end{proof}

\subsection{Three recycled lemmas}

The following lemmas come from  \cite[Lemma 15]{Tomamichel2009}. We need them to prove smoothness of $\Hh^\eps$, so we repeat them here for completeness.

\begin{lemma}\label{lemma:contraction}
 Let $\sigma, \Delta \in \sSet{A}$. The operator
 $$ G :=  \sigma^\half (\sigma + \Delta)^{-\half}$$
 is a contraction, \ie $G \geq 0$ and $\|G\|_\infty \leq 1$.
 In particular, conjugating any positive operator $X$ with $G$ can only decrease the trace:  $ \tr(G\, X\, G^\dag) \leq \tr (X)$.
\end{lemma}
\begin{proof}
We conjugate the following with $(\sigma + \Delta)^{-\half}$,
\begin{align*}
\sigma 
  &\leq \sigma + \Delta \\
(\sigma + \Delta)^{-\half} \sigma (\sigma + \Delta)^{-\half} 
  &\leq  (\sigma + \Delta)^{-\half} (\sigma + \Delta) (\sigma + \Delta)^{-\half}  \\
 G^\dag G &\leq \id.
\end{align*}
Now we use the fact that, for the operator norm, $ \Rightarrow {\|G\|_\infty}^2 = \| G^\dag G \|_\infty \leq \|\id \|_\infty =1$.
The second claim comes from $\tr(G X G^\dag) = \tr(X \,G^\dag\, G) \leq \tr (X \, \id)$.
\end{proof}

\begin{lemma}
 \label{lemma:conjugation}
Let $\rho_{AB} \in \Set{A\otimes B}$, and $\sigma_B, \Delta_B \in \sSet{B} $, such that $\rho_B \leq \sigma_B + \Delta_B$. 
Let $G_B = \sigma^{\half} (\sigma+\Delta)^{- \half}$. 
Then,
$$\| \rho_{AB} -(\id_A \otimes G_B) \rho_{AB} (\id_A \otimes G_B^\dag )\|_1 \leq 2 \sqrt{2 \tr \  \Delta}.$$
\end{lemma}

\begin{proof}
First we work with the fidelity between the two states, and later we relate it to the trace distance. 
Using Uhlmann's theorem, we bound the fidelity using a purification of $\rho_{AB}$. Note that if $\ket \psi_{RAB}$ purifies $\rho_{AB}$, then $(\id_{RA} \otimes G_B) \ket \psi$ purifies 
$(\id_A \otimes G_B) \rho_{AB} (\id \otimes G_B^\dag)$, and in particular it purifies $G_B \rho_B G_B^\dag$.  We have
\begin{align*}
  F(\rho_{AB}, (\id_A \otimes G_B) \rho_{AB} (\id_A \otimes G_B^\dag ))
  &\geq F(\ket \psi, (\id_{RA}  \otimes G_B) \ket \psi) \\
  &= \abs{\bra \psi (\id_{RA} \otimes G_B) \ket \psi} \\
  &=\abs{\tr( (\id_{RA} \otimes G_B) \pure \psi)}  \\
  &=\abs{\tr( G_B \ \rho_B )} \\
  \text{ \flag{real and imaginary parts}} \qquad
  &= \sqrt{\mathcal R[\tr (G_B \ \rho_B)]^2 + \mathcal I[\tr (G_B \ \rho_B)]^2  } \\
  &\geq \mathcal R[\tr (G_B \ \rho_B)] \\
  &= \tr \left( \half (G_B + G_B^\dag) \rho_B \right) .
\end{align*}
From Lemma~\ref{lemma:contraction} we know that $G$ is a contraction. Note that $\half (G + G^\dag)$ is also a contraction, as  $ \| \half (G + G^\dag) \|_\infty \leq \half \| G\|_\infty + \half \|G^\dag\|_\infty \leq 1$. We omit the subscript $B$ in most of the following.
We have
\begin{align*}
  1 - \tr \left( \half (G + G^\dag) \rho_B \right) 
  &= \tr \left( \underbrace{\left[ \id_B -\half (G + G^\dag) \right]}_{\green{\geq0}} \rho_B \right)  \\
  &\leq \tr \left( \left[ \id_B -\half (G + G^\dag) \right] (\sigma + \Delta) \right)  \\
  &= \tr (\sigma+ \Delta) - \half \tr (  (G + G^\dag)  (\sigma + \Delta) )    \\
  &= \tr (\sigma + \Delta) - \half \tr (  \sigma^\half (\sigma+ \Delta)^{-\half} (\sigma + \Delta) ) 
  - \half \tr ((\sigma+ \Delta)^{-\half} \sigma^\half  (\sigma + \Delta)  )  \\
  &=  \tr (\sigma + \Delta) -  \tr (  \sigma^\half (\sigma+ \Delta)^\half ) \\
  &\leq \tr (\sigma+\Delta) - \tr (\sigma) \\
  &= \tr (\Delta),
\end{align*}
so $ F(\rho_{AB}, (\id_A \otimes G_B) \rho_{AB} (\id_A \otimes G_B^\dag ) ) \geq 1 - \tr (\Delta) $. 
From the relation between trace distance and fidelity, we have
\begin{align*}
  \|  \rho_{AB} -(\id_A \otimes G_B) \rho_{AB} (\id_A \otimes G_B^\dag ) \|_1 
   &\leq 2 \sqrt{1 - F(\rho_{AB} , (\id_A \otimes G_B) \rho_{AB} (\id_A \otimes G_B^\dag ) )^2 } \\
   &\leq 2 \sqrt{1 - (1 - \tr (\Delta))^2 } \\
   &= 2 \sqrt{1  -1 + 2 \tr (\Delta) - \tr(\Delta)^2 } \\
   &\leq 2 \sqrt{2 \ \tr (\Delta)}.
\end{align*} 
\end{proof}

The following lemma is simply an adaptation of \cite[Lemma 5.2]{Tomamichel2012} for the alternative smooth min-entropy. The proof is identical. 

\begin{lemma} \label{lemma:direct_sum_state}
 Let $ \rho \in \Set{A\otimes B}, \eps \geq 0$. Then, there is an embedding from $A$ to $A \oplus \bar A$ and a normalized state $\hat \rho \in \Set{(A \oplus \bar A) \otimes B}$ such that 
 \begin{align*}
  \hat H_{\min}^\eps (A|B)_{\rho} = \hat H_{\min} (A \oplus \bar A |B)_{\hat \rho},
 \end{align*}
 with $\hat \rho \in \mathcal B^\eps (\rho)$ (according to the purified distance), and $\abs{\bar A}= \lceil \eps \ 2^{\hat H_{\min}^\eps (A|B)_{\rho}} \rceil$.
\end{lemma}

\begin{proof}
 Let us choose the subnormalized state $\tilde \rho_{AB} \in \sSet{A\otimes B}$ that achieves the maximum in the definition of the entropy, i.e., 
 \begin{align*}
  &\lambda = \hat H_{\min}^\eps (A|B)_{\rho}, \\
  &\tilde \rho_{AB} \leq 2^{-\lambda} \ \id_A \otimes \tilde \rho_B.
 \end{align*}
 Now we construct the direct sum space $A \oplus \bar A$, where $\bar A$ is a Hilbert space of dimension $\abs{\bar A} \geq \eps \ 2^\lambda $. 
 In that space, we write a normalized extension of $\tilde \rho$,
 \begin{align*}
  \hat \rho = \tilde \rho_{AB} \oplus \left( (1 - \tr \tilde \rho) \frac{\id_{\bar A}}{\abs{\bar A}} \otimes \tilde \rho_B \right) \quad
  \in \Set{(A \oplus \bar A) \otimes B} .
 \end{align*}
Note that $\hat \rho_B \propto \tilde \rho_B$.
We have 
\begin{align*}
 \hat \rho 
   &= \underbrace{\tilde \rho_{AB}}_{\green{\leq  2^{-\lambda} \ \id_A \otimes \tilde \rho_B }} \oplus \left( \underbrace{(1 - \tr \tilde \rho)}_{\green{\leq \eps}} \frac{\id_{\bar A}}{\abs{\bar A}} \otimes \tilde \rho_B \right) \\
   &\leq (2^{-\lambda}\ \id_A  \otimes \tilde \rho_B ) \oplus \left( \frac{ \eps}{\eps \ 2^\lambda } \id_{\bar A} \otimes \tilde \rho_B \right)  \\
   &= 2^{-\lambda} (\id_A \oplus \id_{\bar A})  \otimes \tilde \rho_B.
\end{align*}
This tells us that $\lambda$ is a feasible candidate for the primal SDP of the entropy of $\hat \rho$, and therefore
\begin{align*}
 \hat H_{\min} (A \oplus \bar A | B)_{\hat \rho} \geq \lambda = \hat H_{\min}^\eps (A  | B)_\rho.
\end{align*}
Now we show that $\hat \rho \in \mathcal B^\eps (\rho)$, according to the purified distance. It suffices to show that $F(\rho, \hat \rho) = F(\rho, \tilde \rho)$.  
The fidelity is linear under direct sums, i.e., for two states  $\sigma = \sigma_1 \oplus \sigma_2 $ and $\tau = \tau_1 \oplus \tau_2 $, we have 
\begin{align*}
F(\sigma, \tau) &= \| \sqrt \sigma \sqrt \tau \|_1 
  = \| \sqrt {\sigma_1 }\sqrt {\tau_1} \|_1  + \| \sqrt {\sigma_2} \sqrt {\tau_2} \|_1 .
\end{align*}
In our case, we have $\rho = \rho_{AB} \oplus 0_{\bar A B}$ and $\hat \rho = \tilde \rho_{AB} \oplus \left( (1 - \tr \tilde \rho) \frac{\id_{\bar A}}{\abs{\bar A}} \otimes \tilde \rho_B \right) $, so
\begin{align*}
 F(\rho, \hat \rho) 
   &= \|\sqrt \rho \sqrt {\hat \rho} \|_1 \\
   &= \| \sqrt{ \rho_{AB} } \sqrt{\tilde \rho_{AB}} \|_1 + \|  \sqrt{0_{\bar A B}} \sqrt {\left( (1 - \tr \tilde \rho) \frac{\id_{\bar A}}{\abs{\bar A}} \otimes \tilde \rho_B \right)} \|_1  \\
   &= F(\rho, \tilde \rho) + 0,   
\end{align*}
which implies $\hat \rho \in \mathcal B^\eps (\rho)$. Therefore we have, by definition of the smooth entropy,
\begin{align*}
 \hat H_{\min} (A \oplus \bar A | B)_{\hat \rho} \leq  \hat H_{\min}^\eps (A | B)_\rho,
\end{align*}
and the equality follows.  
\end{proof}

\subsection{Smoothness of $\Hh^\eps$ and relation to $\hmin^{\eps'}$}

The next lemma proves that the generalized smooth entropy is actually ``smooth'', i.e., 
if two states $\rho$ and $\sigma$ are close according to the trace distance, then their generalized smooth entropies are also close. 

\begin{lemma}\label{lemma:smoothness_Hh}
Let $\rho_{AB}, \sigma_{AB} \in \Set{A\otimes B}$ be two positive, normalized density operators, such that $\| \rho_{AB} - \sigma_{AB} \|_1 \leq \delta$.
Then, for any $\eps > 0$,
\begin{align*}
 \Hh^{\eps}(A|B)_\rho
 \leq  \Hh^{\eps+\delta + 2 \sqrt{2\delta}}(A|B)_\sigma
    + \log \frac{\eps + \delta + 2\sqrt{2\delta}}\eps.  
\end{align*}
\end{lemma}

\begin{proof}

This proof is made of two parts. First we will relate $\Hh^{\eps}(A|B)_\rho$ to $-D^\eps(\rho_{AB} - \Delta' || \id_A \otimes \sigma_B )$, where  $\Delta'$ is a positive operator with trace at most $2 \sqrt{2\delta}$. Later we bound $-D^\eps(\rho_{AB} - \Delta' || \id_A \otimes \sigma_B )$ in terms of $\Hh^{\eps+\delta + 2 \sqrt{2\delta}}(A|B)_\sigma$.

We have 
\begin{align*}
  \|\rho_B - \sigma_B \|_1  
  \leq \|\rho_{AB} - \sigma_{AB} \|_1   
  \leq \delta ,
\end{align*}
and therefore there exist positive operators $\Delta^+  $ and $ \Delta^-$ such that
\begin{align*}
 \rho_B - \sigma_B 
   &= \Delta^+  - \Delta^-,  \qquad
   \Delta^+ , \Delta^- \geq 0, \qquad
   \tr( \Delta^+ ), \tr( \Delta^-) \leq \delta. \\
 \rho_B &\leq \sigma_B + \Delta^+ . 
\end{align*}

Consider the pair $(\mu, X)$ that forms the optimal solution of the dual SDP for $H^\eps(A|B)_\rho$,
\begin{align*}
 2^{H^\eps(A|B)_\rho} = \mu - \tr \frac X \eps, \qquad
 \mu\ \rho_{AB} &\leq \id_A \otimes \rho_B + X_{AB} .
\end{align*}
We can define the operator 
\begin{align*}
  G = \sigma_B^\half (\sigma_B + \Delta^+)^{-\half}.
\end{align*}
We conjugate the feasibility condition for the dual program with $\id_A \otimes G)$, 
\begin{align*}
\mu\ \rho_{AB} &\leq \id_A \otimes \rho_B + X_{AB} \\
\mu\ (\id \otimes G) \ \rho_{AB} \ (\id \otimes G^\dag ) &\leq   \id_A \otimes G \ \rho_B \ G^\dag + (\id \otimes G ) X_{AB} (\id \otimes G^\dag ). 
\end{align*}
On the right-hand side, we have
\begin{align*}
  G \ \rho_B \ G^\dag 
  &\leq G \ (\sigma_B + \Delta^+)  G^\dag  \\
  &= \sigma_B^\half (\sigma_B + \Delta^+)^{-\half} (\sigma_B + \Delta^+) (\sigma_B + \Delta^+)^{-\half} \sigma_B^\half \\
  &= \sigma_B. 
\end{align*}
Note also that $G$ is a contraction (see Lemma~\ref{lemma:contraction}, and therefore $\tr ((\id \otimes G ) X_{AB} (\id \otimes G^\dag )) \leq \tr (X_{AB})$.
On the left-hand side, we apply Lemma~\ref{lemma:conjugation}, 
\begin{align*}
 &\| \rho_{AB} -(\id_A \otimes G_B) \rho_{AB} (\id_A \otimes G_B^\dag )\|_1 \leq 2 \sqrt{2 \tr \Delta^+} \\
 \Rightarrow \quad \exists \Delta': \quad
 &\rho_{AB} - \Delta' 
 \leq
 (\id_A \otimes G_B) \rho_{AB} (\id_A \otimes G_B^\dag ) , 
 \qquad
 \Delta' \geq 0, \  \tr (\Delta') \leq 2 \sqrt{2 \tr \Delta^+} \leq   2 \sqrt{2 \delta} .
\end{align*}
This gives us 
\begin{align*}
  \mu\ (\rho_{AB} - \Delta') 
   &\leq   \id_A \otimes \sigma_B 
      + \underbrace{  (\id \otimes G ) X_{AB} (\id \otimes G^\dag ) }_{\green{=: X' \geq 0}} 
\end{align*}
The above inequality tells us that $(\mu, X')$ form a candidate pair for the dual SDP of $D^\eps (\rho_{AB} - \Delta ' || \id_A \otimes \sigma_B )$. 
Note that $\rho_{AB}-\Delta'$ is Hermitian (as both $\rho_{AB}$ and $\Delta'$ are positive operators, and therefore Hermitian), so both primal and dual SDPs for $D^\eps(\rho_{AB} - \Delta' || \id_A \otimes \sigma_B )$ are well defined. 
Since the dual program is a maximization over all feasible pairs, we have
\begin{align*}
  2^{-D^\eps(\rho_{AB} - \Delta' || \id_A \otimes \sigma_B )} 
  &\geq \mu - \frac{\tr \ X'}\eps    \\
  &=\mu - \frac{\tr [  (\id \otimes G )  X_{AB} (\id \otimes G^\dag )   ]}\eps  \\
  &\geq \mu - \frac{\tr (X_{AB})  }\eps \\ 
  &= 2^{H^\eps (A|B)_\rho}.
\end{align*}
This gives us the bound
\begin{align}
  -D^\eps(\rho_{AB} - \Delta' || \id_A \otimes \sigma_B )
  &\geq H^\eps (A|B)_\rho  .
  \label{eq:Dh_Hrho}
\end{align}
Now we just need to relate $ D^\eps(\rho_{AB} - \Delta' || \id_A \otimes \sigma_B ) $ to  the smooth conditional entropy of $\sigma$, $H^{\eps'}(A|B)_\sigma$ (which, as we will see, might have a different smoothing parameter, $\eps'$). 
First we observe that the operator $\rho_{AB} - \Delta'$ is close to $\sigma$, 
\begin{align*}
 \| (\rho_{AB} - \Delta') - \sigma_{AB}\|_1 
  &\leq \delta + \tr \Delta' 
  \nonumber \\
  &\leq \underbrace{ \delta + 2 \sqrt{2 \delta}}_{\green{=: \delta'}}.
\end{align*}
To shorten notation, it is convenient to define $\delta' := \delta + 2 \sqrt{2 \delta}$.
The trace distance gives us an upper bound for the probablity of distinguishing two states by applying any POVM $\set{Q, \id-Q}$, \
\begin{align}
 \max_{0\leq Q\leq \id} \abs{ \tr (Q \ [\rho_{AB} - \Delta'] ) - \tr (Q\ \sigma_{AB}) } &\leq \delta'.
   \label{eq:close_Q}
\end{align}
We start by writing down the primal SDP for
$2^{H^{\eps+\delta'} (A|B)_\sigma}$,

\begin{center}
\begin{tabular}{r  l}
 minimize \quad
   & $\frac1{\eps +\delta'} \tr (Q \  \id_A \otimes \sigma_B  )$ \\
 \quad \\
 subject to \quad
   & $\tr(Q \ \sigma_{AB}) \geq \eps +\delta' ,$\\
   &$0 \leq Q \leq \id$.
\end{tabular} 
\end{center}

We take the operator $Q$ that achieves the minimum, and show that $Q$ is a feasible candidate for the primal SDP of $2^{-D^{\eps}(\rho_{AB} - \Delta' || \id_A \otimes \sigma_B )}$. To make that clear, let us first write this SDP, 

\begin{center}
\begin{tabular}{r  l}
 minimize \quad
   & $\frac1{\eps } \tr (P \  \id_A \otimes \sigma_B  )$ \\
 \quad \\
 subject to \quad
   & $\tr(P \ [\rho_{AB} - \Delta']) \geq \eps ,$\\
   &$0 \leq P \leq \id$.
\end{tabular} 
\end{center}
We may relate the feasibility conditions of the two SDPs using inequality~\ref{eq:close_Q}, which gives us
\begin{align*}
  \tr (Q\ [\rho_{AB}- \Delta']) 
  &\geq  \tr (Q \  \sigma_{AB} ) - \delta' \\
  &\geq \eps + \delta' - \delta' = \eps.
\end{align*}
Therefore we can  bound $2^{-D^\eps(\rho_{AB} - \Delta' || \id_A \otimes \sigma_B )}$ as
\begin{align*}
 2^{-D^{\eps}(\rho_{AB} - \Delta' || \id_A \otimes \sigma_B )}
    &\leq \frac1\eps  \tr (Q \ \id_A\otimes \sigma_B) \\
    &= \frac1\eps  (\eps + \delta') 2^{\Hh^{\eps+\delta'}(A|B)_\sigma} .
\end{align*}
Taking the logarithm and using $\delta'= \delta + 2 \sqrt{2\delta}$, we obtain
\begin{align*}
 \Hh^{\eps+\delta + 2 \sqrt{2\delta}}(A|B)_\sigma
 &\geq - \Dh^{\eps }(\rho_{AB}- \Delta'||\id_A \otimes \sigma_B) 
    - \log \frac{\eps + \delta + 2 \sqrt{2\delta}}\eps \\
    \green{\text{(\eref{eq:Dh_Hrho})} }  \qquad 
 &\geq \Hh^{\eps}(A|B)_\rho
  - \log \frac{\eps + \delta + 2 \sqrt{2\delta}}\eps.  
\end{align*}
\end{proof}

% xxxxxxxxxxxxxxxxxxxxxx smooth min-entropy xxxxxxxxxxxxxxxxxx

In the following lemma, we find a lower bound for $H^\eps$ in terms of the smooth min-entropy. An upper bound is given in \cite[Prop.~4.1]{Dupuis2012}.

\begin{lemma}\label{lemma:min_and_Hh}
Let $\rho \in \Set{A\otimes B}$, and let $\eps \in ]0, \frac12]$ . Then,
\begin{align*}
  \hmin^\eps (A|B)_\rho 
 \leq  \Hh^{11\sqrt{\eps}} (A |B)_\rho
   - \frac52  \log \left(\frac3\eps  \right)
    + \log \left( \frac2{1-\eps}\right).
\end{align*}

\end{lemma}

\begin{proof}
 
 See \fref{fig:hell} for a schematic representation of the different steps of this proof.
 
 \begin{figure}[h]
   \centering
   \includegraphics[width=0.6 \textwidth]{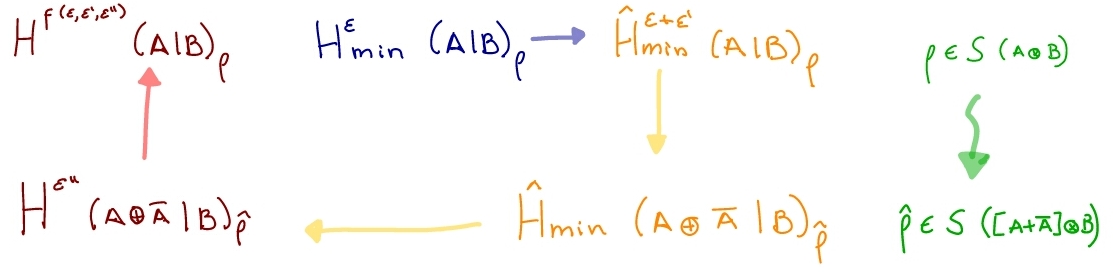}
   \caption{Diagram illustrating all the steps of the proof of Lemma~\ref{lemma:min_and_Hh}. We tried to make it less terrifying by using friendly colours. We start from $\hmin^\eps (A|B)_\rho$ and bound it successively until we end up with the generalized smooth entropy for the same state. Along the way we need to extend our state $\rho$ to a larger Hilbert space (below).}
   \label{fig:hell}
\end{figure}

 From \cite[Lemma 19]{Tomamichel2011} we have that
 \begin{align*}
  \forall \rho \in \Set{A\otimes B}, \ \forall \eps,  \eps' \in\ ]0,1], \quad 
  \hmin^\eps (A|B)_\rho \leq  \hat H_{\min}^{\eps + \eps'} (A|B)_{\rho} + \log \left( \frac2{{\eps'}^2} + \frac1{1-\eps} \right),
 \end{align*}
  Now we use Lemma~\ref{lemma:direct_sum_state} to find a normalized state $\hat \rho \in \mathcal B^{\eps+\eps'} (\rho)$ in a larger Hilbert space  $(A \oplus \bar A) \otimes B$ that attains the optimization. This gives us
 \begin{align*}
 \hmin^\eps (A|B)_\rho 
    &\leq  \hat H_{\min}^{\eps + \eps'} (A|B)_{\rho} + \log \left( \frac2{{\eps'}^2} + \frac1{1-\eps} \right)   \\
    &= \hat H_{\min} (A \oplus \bar A|B)_{\hat \rho} + \log \left( \frac2{{\eps'}^2} + \frac1{1-\eps} \right).
 \end{align*}
It follows from the definition  of $\hat H_{\min}$ that \cite[Prop.~4.1]{Dupuis2012}
\begin{align*}
   \forall \ \eps'' \in\ ]0,1]: \quad \hat H_{\min} (A \oplus \bar A|B)_{\hat \rho}  \leq \Hh^{\eps ''} (A \oplus \bar A|B)_{\hat \rho} ,
\end{align*}
which leaves us with 
 \begin{align*}
 \hmin^\eps (A|B)_\rho 
    &\leq  \Hh^{\eps ''} (A \oplus \bar A|B)_{\hat \rho} + \log \left( \frac2{{\eps'}^2} + \frac1{1-\eps} \right) .
 \end{align*}
 Now we only need to relate  $\Hh^{\eps ''} (A \oplus \bar A|B)_{\hat \rho} $ back to the smooth entropy of $\rho$. 
 Since the two states $\rho, \hat \rho$ are normalized, we have $\| \rho - \hat \rho\|_1 \leq 2 (\eps + \eps')$. We can use Lemma~\ref{lemma:smoothness_Hh} to obtain
 \begin{align*}
 \Hh^{\eps''}(A \oplus \bar A|B)_{\hat\rho}
 \leq  \Hh^{\eps'' + 2 \eps +2\eps' + 4 \sqrt{\eps + \eps'}}(A\oplus \bar A |B)_\rho
    + \log \frac{\eps'' + 2 \eps +2\eps' + 4 \sqrt{\eps + \eps'} }{\eps''}.  
\end{align*}
Now we observe that $\rho$ has no support on $\bar A$, therefore, for any smoothing factor $\tilde\eps \in [0,1]$,
\begin{align*}
 \Hh^{\tilde \eps}(A\oplus \bar A |B)_\rho = \Hh^{\tilde \eps}(A|B)_\rho.
\end{align*}
All in all, we have
\begin{align*}
 \hmin^\eps (A|B)_\rho 
 \leq  \Hh^{\eps'' + 2 \eps +2\eps' + 4 \sqrt{\eps + \eps'}}(A |B)_\rho
    + \log \left( \frac{\eps'' + 2 \eps +2\eps' + 4 \sqrt{\eps + \eps'} }{\eps''} \right)
     + \log \left( \frac2{{\eps'}^2} + \frac1{1-\eps} \right).
\end{align*}
To clean up,  we consider the special case $\eps = \eps'= \eps''$, which gives us
\begin{align*}
  \hmin^\eps (A|B)_\rho 
 \leq  \Hh^{5 \eps + 4 \sqrt{2 \ \eps}} (A |B)_\rho
    + \log \left(5 +  \frac{4 \sqrt 2  }{\sqrt \eps} \right)
     + \log \left( \frac2{{\eps}^2} + \frac1{1-\eps} \right),
\end{align*}
and finally we upper bound the additive terms and smoothing factors with simpler terms (the factors were found numerically). We obtain
\begin{align*}
  \hmin^\eps (A|B)_\rho 
 \leq  \Hh^{11\sqrt{\eps}} (A |B)_\rho
    - \frac52  \log \left(\frac\eps3   \right) 
    + \log \left( \frac2{1-\eps}\right).
\end{align*}

\end{proof}

%............. Bibliography ..............

\end{document}